\documentclass[iop]{emulateapj}

\shorttitle{REXCESS UV}
\shortauthors{Donahue et al.}
\usepackage{graphics,graphicx,epsf,natbib}
\received{}

\begin{document}
\title{Star Formation and UV Colors of the Brightest Cluster Galaxies \\ 
in the Representative XMM-Newton Cluster Structure Survey}
\author{
Megan Donahue\altaffilmark{1}, Seth Bruch\altaffilmark{1},  Emily Wang\altaffilmark{1}, G. Mark Voit\altaffilmark{1},
Amalia K. Hicks\altaffilmark{1}, Deborah B. Haarsma\altaffilmark{2},  Judith H. Croston\altaffilmark{3}, Gabriel W. Pratt\altaffilmark{4}, Daniele Pierini\altaffilmark{5}, Robert W. O'Connell\altaffilmark{6}, Hans B{\"o}hringer\altaffilmark{5}}

\altaffiltext{1}{Michigan State University, Physics \& Astronomy Dept., East Lansing, MI 48824-2320, donahue@pa.msu.edu}
\altaffiltext{2}{Calvin College, 1734 Knollcrest Circle SE, Grand Rapids, MI 48546}
\altaffiltext{3}{School of Physics and Astronomy, University of Southampton, Southampton, SO17 1SJ, UK}
\altaffiltext{4}{Laboratoire AIM, IRFU/Service d’Astrophysique - CEA/DSM - CNRS - Universit{\'e} Paris Diderot, Bât. 709, CEA-Saclay, F-91191, Gif-sur- Yvette Cedex, France}
\altaffiltext{5}{Max-Planck-Institut f\"ur extraterrestrische Physik, Giessenbachstr., 85748 Garching, Germany}
\altaffiltext{6}{Dept. of Astronomy, University of Virginia, Charlottesville, VA 22904-4325}

\begin{abstract}
We present UV broadband photometry and optical emission-line measurements for a sample of 32 Brightest 
Cluster Galaxies (BCGs) in clusters of the Representative XMM-Newton Cluster Structure Survey (REXCESS) with $z=0.06-0.18$. 
The REXCESS clusters, chosen to study scaling relations in clusters of galaxies, have X-ray measurements of high quality. 
The trends of star formation and BCG colors with BCG and host properties can be investigated with this sample.
The UV photometry comes from the XMM Optical Monitor, supplemented by existing archival 
GALEX photometry. We detected  H$\alpha$ and forbidden line emission in 7 (22\%) of these BCGs, in optical spectra obtained using the
SOAR Goodman Spectrograph. All of these emission-line BCGs occupy  
clusters classified as cool cores based on the central cooling time in the cluster core, for
an emission-line incidence rate of 70\% for BCGs in REXCESS cool core clusters. 
Significant correlations between the H$\alpha$ equivalent widths, excess UV production in the BCG, and 
the presence of dense, X-ray bright intracluster gas with a short cooling time are seen, including
the fact that all of the H$\alpha$ emitters inhabit systems with short central
cooling times and high central ICM densities. Estimates of the star formation rates based on 
H$\alpha$ and UV excesses are consistent with each other in these 7 systems, ranging from $0.1-8$ solar masses per year.
The incidence of emission-line BCGs in the REXCESS sample is
intermediate, somewhat lower than in other X-ray selected samples ($\sim35\%$), and somewhat higher than but
statistically consistent with optically selected, slightly lower redshift BCG samples ($\sim10-15\%$).
The UV-optical colors (UVW1-R $\sim 4.7 \pm 0.3$)  of REXCESS BCGs without strong optical 
emission lines are consistent with those predicted  from templates and observations of ellipticals 
dominated by old stellar populations.  
We see no trend in UV-optical colors with optical luminosity, $R-K$ color, X-ray temperature, redshift, or
offset between X-ray centroid and X-ray peak ($\langle w \rangle$). The lack of such trends in these massive galaxies, particularly the ones lacking
emission lines, suggests that the proportion 
of UV-emitting (200-300 nm) stars is insensitive to galaxy mass, cluster mass, cluster
relaxation, and recent evolution, over the range of this sample.

\end{abstract}

\keywords{galaxies:clusters:general, cooling flows, galaxies: elliptical and lenticular, cD}

\section{Introduction}

In current models of hierarchical galaxy formation, 
Brightest Cluster Galaxies (BCGs) are the trash heaps of the universe. As galaxies fall through and 
past the center of a cluster of galaxies, tidal forces strip them of their stars, and ram pressure
stripping by the dense intracluster medium (ICM) removes some of their gas. The BCG settles to the cluster's center
of mass and accretes stars and gas, developing a huge 
extended stellar halo characteristic of central dominant (cD) galaxies.
These galaxies are therefore not simply overgrown massive ellipticals, but represent a
category of galaxies that appear to have very special growth histories.

Modern simulations have great difficulty 
reproducing the observed properties of these singular objects, which form deep in the potential well of a cluster. 
Apparently, the star formation in these systems must be quenched in order to explain their optical luminosities
and colors \citep[e.g., ][]{1998MNRAS.294..705K}. In the most massive galaxies, the primary agent of feedback at late times 
is thought to be the central supermassive black hole.  The vast majority of the stars in the most massive elliptical galaxies were already present billions of years ago, based on studies of their color-magnitude relation and spectral energy distributions \citep[e.g., ][]{1992MNRAS.254..601B,1998A&A...334...99K, 2008MNRAS.386.1045A}, and the mass of a galaxy's central black hole tends to be about 0.2\% the mass of its spheroidal component \citep[e.g., ][]{1998AJ....115.2285M,2000ApJ...539L..13G,2000ApJ...539L...9F}. Somehow, formation and growth of the black hole is coupled to the formation and growth of the galaxy.  Meanwhile, the downsizing phenomenon 
\citep{1999AJ....118..603C}, wherein massive galaxies stop forming stars earlier than low-mass galaxies, suggests that star formation somehow decouples from the growth of the galaxy by hierarchical accretion \citep[e.g., ][]{2002MNRAS.333..156B}.  Galaxy formation models without feedback from an active galactic nucleus (AGN) predict high-mass galaxies that are far too blue and luminous \citep{1998MNRAS.294..705K}, but the situation may be resolved when (sufficient) feedback from an AGN is included \citep[e.g., ][]{1997ApJ...487L.105C, 1998A&A...331L...1S, 2004MNRAS.347.1093B}. 

The star formation in today's BCGs, if present at all, is a mere shadow of what it must have been 
billions of years ago when most of the stars were formed.  Nevertheless, some BCGs apparently do persist
in forming stars, with the trend that the clusters with the shortest cooling times appear
to be much more likely to host emission line systems and blue cores \citep[e.g., ][]{1983ApJ...272...29C, 1989ApJ...338...48H, 1992ApJ...393..579M,2006ApJ...652..216R,2008ApJ...683L.107C, 2008ApJ...681L...5V}. Such 
clusters comprise about half of X-ray luminous 
($L_X \gtrsim 10^{44}$ erg s$^{-1}$)
clusters at low redshift \citep{1992MNRAS.258..177E,1992ApJ...385...49D,1999MNRAS.306..857C}.
\citet[][]{2005ApJ...635L...9H}, in an XMM Optical Monitor study of 33 galaxies, of which
9 were BCGs, showed a connection between excess ultraviolet (UV) emission and recent X-ray estimates of mass cooling rates. There appears to be 
a clear empirical connection between the state of the hot gas and the activity in the BCG. 

Observationally, multi-wavelength studies of clusters of galaxies and their BCGs provide some
insight into how this feedback process may proceed. Cavities in the X-ray intracluster gas that surrounds  
radio lobes provided smoking-gun evidence for AGN feedback
\citep[e.g., ][]{1993MNRAS.264L..25B,2005Natur.433...45M}. Estimates of the $PdV$ work required to inflate such cavities have shown that the kinetic energy outputs from these AGNs are surprisingly high compared to their radio luminosities and estimated lifetimes \citep[e.g., ][]{2007ARA&A..45..117M}. Central radio sources are quite common at the centers of galaxy clusters with cool, dense ICM cores \citep{1990AJ.....99...14B}.  
Sporadic AGN outbursts with kinetic outputs of $\sim 10^{45} \, {\rm erg \, s^{-1}}$ can plausibly stabilize cooling and star formation in the BCGs of those clusters, explaining why those galaxies are less red and less luminous than BCGs modeled without AGN feedback \citep[e.g., ][]{2005ApJ...634..955V}. While this is not necessarily the feedback mode that quenches star formation in elliptical galaxies at high redshift, this is the only form of AGN feedback that we can currently study in such detail.

Multi-wavelength studies of a well-chosen X-ray sample without any particular morphological criteria, with uniform
and accurate X-ray measurements,  
are the best way to study the relationship of star formation, intracluster gas, and AGN feedback at low redshift. Such studies 
shed light on what regulates star formation in large galaxies at earlier times in the universe.
Ultraviolet observations are sensitive to the continua of OB stars, and H$\alpha$ is produced
by photoionization of the interstellar medium (ISM) by O-stars.  Elliptical galaxies, and therefore older stellar populations, 
exhibit an increasing UV emission towards shorter UV wavelengths (``UV upturn'')  
from extreme horizontal branch stars.  However, recent discussions
suggest that relatively tiny amounts of young stars can contribute to the observed scatter in 
this component for UV starlight \citep[e.g., ][]{2007MNRAS.380.1098H}. 

In this work we investigate the UV properties of BCGs including both UV from recent star formation
and from older stellar populations.
We present new UV observations from the {\em XMM-Newton} Optical Monitor and archival results from the
Galaxy Evolution Explorer (GALEX), 
together with  a census of BCG emission-line activity from ground-based observations from the SOAR Goodman Spectrograph.
These observations are of BCGs in the Representative {\em XMM--Newton} Cluster Structure Survey 
(REXCESS) \citep{2007A&A...469..363B}. We provide a small number of 
supplemental observations to complete the spectroscopic coverage of a comparison sample of BCGs.
We assume cosmological parameters $H_0=70$ km s$^{-1}$ Mpc$^{-1}$ ($h=0.7$) 
and a flat geometry with $\Omega_M=0.3$ throughout.

\section{Sample Description}
The Representative {\em XMM-Newton} Cluster Structure Survey 
(REXCESS) \citep{2007A&A...469..363B} was designed to investigate how the properties of galaxy clusters scale with mass. 
Its primary scientific goal was to further the understanding of how cluster observables such as luminosity 
and temperature depend on cluster mass, in order to improve the calibration of the scaling 
relations essential for cluster cosmology and to provide a
comparison sample for galaxy cluster simulations. The sample was selected on the basis of X-ray
luminosity alone, with no bias regarding morphology or central surface brightness. It was also
selected to span a wide range in mass, luminosity, and X-ray temperature ($T_X=2-9$ keV). 
The sample was chosen to optimally cover the full range of cluster X-ray luminosities, at
redshifts such that the cluster is compact enough for local background estimation with XMM 
yet extended enough for measuring temperature and surface-brightness structure ($z\sim0.07-0.18$).

The REXCESS sample differs in important respects from that of a previous XMM Optical Monitor study 
\citep{2005ApJ...635L...9H}. The Hicks \& Mushotzky sample was smaller (only 9 BCGs) and 
heterogeneous, chosen from
the literature for the existence of published optical and X-ray properties and for their proximity,
with a few added at moderate redshift for their extreme cool core properties. In contrast, the REXCESS
sample is specifically designed to be representative for a given cluster luminosity range, 
which means that we can now ask whether the UV colors, star formation rate (as sampled by H$\alpha$
or a UV excess over that of an old population), or incidence of star formation signatures
scale with cluster luminosity, cluster mass, or BCG stellar mass. 

Ten of these REXCESS BCGs reside in clusters classified as cool cores on 
the basis of their short central cooling times ($< 1.7$ Gyr) and large central ICM electron 
densities \citep{2009A&A...498..361P}.  Among the 10 BCGs in cool-core clusters, 8 have powerful radio luminosities ($>10^{24}$ W Hz$^{-1}$ 
at 1.4 GHz). These estimates are based on NVSS fluxes and converted 843 MHz fluxes from SUMSS, assuming a spectral power law index 
of -0.7 (Heidenreich et al., in prep.), in accord with the finding that radio-loud 
BCGs are found only at the centers of cool-core clusters or in galaxy-scale ($<4$ kpc) 
coronae \citep{2008ApJ...683L.107C,2008MNRAS.385..757D,2009ApJ...704.1586S}. 
One non-CC cluster, RXCJ0211.4-4017, is on the border between CC and non-CC based on its cooling time, also
has a similarly powerful radio source, a source which should be in a compact X-ray corona,
based on the predictions by \citet{2009ApJ...704.1586S}.

\citet{2009A&A...498..361P} present the key X-ray 
scaling relations between $L_X$, $T_X$, $Y_X$ (product of gas mass 
and $T_X$), and $M$ among the REXCESS clusters. The measurement errors are negligible ($1-3\%$) compared
to the intrinsic scatter. They found that the main source of scatter in the $L_X-T_X$ relation and
other scaling relations is the variation of gas
fraction with mass. Cool core (CC) clusters, defined by their central gas density and
cooling time \citep{2008A&A...487..431C}, occupy the high-luminosity envelope of the relations
while morphologically disturbed (D) clusters, defined by size of their centroid shift, occupy the
low-luminosity envelope. While exclusion of the central region significantly reduces the scatter
in these relations,  (see also \citet{1998ApJ...504...27M, 1994MNRAS.267..779F}),  
the segregation of CC and D systems persists. 
We show here that the CC and non-CC in REXCESS host distinctly different populations of BCGs, 
confirming trends seen in optical emission lines \citep[e.g., ][]{1983ApJ...272...29C, 1989ApJ...338...48H}
and optical colors \citep[e.g., ][]{1989AJ.....98.2018M}.

Mergers may play an important role  in setting
the central entropy level of the non-CC systems in REXCESS \citep{2010A&A...511A..85P}.
\citet{2010ApJ...713...1037H} present R-band photometry for the BCGs in REXCESS clusters 
and analyze it in context with the cluster X-ray properties \citep{2009A&A...498..361P,2008A&A...487..431C}.  
The most intriguing correlation was between the central ICM electron density and
the optical light in a metric $r=12 h^{-1}$ kpc aperture, but only for non-CC clusters. They suggest that
this correlation may be based on common physical processes - mergers and interactions - 
setting the density of stars and the density of gas in cluster cores with long cooling times.
We use the same optical images and surface brightness profiles as in \citet{2010ApJ...713...1037H}, 
although we adopt a slightly smaller metric aperture ($r=10 h^{-1}$ kpc) than used in
that work. 

\section{Ultraviolet Photometry}

The Optical Monitor (OM) Telescope is a 30 cm optical/UV telescope co-aligned with the X-ray telescopes
on board {\em XMM- Newton} \citep{2001A&A...365L..36M}. It is capable of simultaneous observations with the XMM X-ray detectors.
We present data taken with the UVW1 (270 nm) and the UVM2 (220 nm) filters. The 
detector is a micro-channel plate intensified photon-counting CCD. The field of view is $17\arcmin$ across, 
$256 \times 256$ instrument pixels. Photons are centroided to within 1/8 of an instrument pixel, 
yielding $2048 \times 2048$ effective centroiding pixels, each $0.4765\arcsec$ on a side, 
which are binned $2\times2$ in the low resolution images (LR) and are unbinned in the high resolution (HR) 
images. Because of  limitations on on-board memory, the LR images are limited to $488 \times 488$ binned pixels, and
HR images are up to $652 \times 652$ unbinned pixels. The OM observations used for photometry
are listed in Table~\ref{table:om}.

In the LR-mode, 5 exposures are required to cover the full field of view. Therefore, 
we used public, mosaicked OM images created for each observation (ObsID) by \citet{2008PASP..120..740K}.   
Since coincidence-loss and dead-time corrections are negligible for faint sources, photometry of faint sources is possible 
with these mosaics \citep{2008PASP..120..740K}. We also required upper limits for non-detections, photometry for 
newer data, and photometry in metric apertures, so we did our own photometry of the mosaics. 
 For a handful of sources, we compared the photometry from
individual frames with that obtained from the mosaic, and the results were consistent.
Our photometry of well-detected, compact sources spatially coincident
with those published in the OMCAT \citep{2008PASP..120..740K}, as well as
photometry in the XMM-OM SUSS\footnote{The Serendipitous
Ultra-violet Source Survey (\url{http://heasarc.gsfc.nasa.gov/W3Browse/all/xmmomsuss.html}) was created by 
the University College London's Mullard Space Science Laboratory on behalf of ESA.} gave similar results. Some  
discrepancies were found for extended sources, which was not surprising, since the catalogs are
optimized for point source photometry.

Each mosaic frame was normalized to an effective exposure of 1000 seconds. The actual exposure time in
the regions covered by the aperture was derived
from the exposure maps accompanying each mosaic, in order to reconstruct
the true number of detected counts for estimates of statistical uncertainty. 
The images are marred by slightly elevated 
count rates in the center of the field from scattered light.  
The BCG was identified from R-band imaging to be the brightest cluster galaxy within
$500 h^{-1}$ kpc of the X-ray centroid; in two instances (RXCJ2234.5-3744 and RXCJ1311.4-0120) this was
not a clearcut choice, and those are discussed in \citet{2010ApJ...713...1037H}.   We have included the 2MASS identifier
for the BCG we identified for each cluster. 

We used circular apertures with a radius of $10h^{-1}$ kpc, where $h=0.70$, in order to derive a 
UV-optical (R) color. The full width, half maximum (FWHM) of the point spread function (PSF) of the Optical Monitor in the UV is typically
$\sim 1.8-2\arcsec$ for the UVM2 and UVW1 filters, respectively.
The aperture radii, chosen for ease of comparison with BCG photometry in the literature
\citep[e.g.][]{1995ApJ...440...28P}, are over double the FWHM of the PSF even for the most distant
sources in our sample. We masked contaminating UV foreground or background sources in source and background apertures.  
We measured the background count rate in an annulus around the source. We expected scattered
light to be a factor in background estimates, but the galaxies were sufficiently compact compared
to the gradient induced by scattered light that this light proved to be mainly a cosmetic nuisance.
We also obtained a separate estimate of the background count rate in multiple apertures similar to 
the source aperture. A comparison of the histogram of aperture count rates indicated that the assumption of Poisson
statistics in the background was good if we avoided areas of scattered light, but the local annulus
provided the most reliable estimate of the background near the source. 
The count rates were then
converted to AB magnitudes using the zeropoints from the XMM Data Handbook: 
18.5662 and 17.4120 for UVW1 and UVM2, respectively.
All colors plotted in this paper have been corrected for Galactic extinction using the $A_V$ listed in Table~\ref{table:omphot} and
the attenuation model from \citet{2003ApJ...599L..21F,2005ApJ...619..340F}.
For reference, the color term subtracted from the apparent magnitude was $f A_V$, where $f$
was 2.11 for the UVW1 filter, 3.25 for UVM2 and the GALEX NUV filters, and 2.5 for the GALEX
FUV filter. This model gives optical attenuation estimates consistent with those derived from the infrared maps of the
Galaxy \citep{1998ApJ...500..525S}.

The signal-to-noise of each source was estimated from the relation:
\begin{equation}
SNR = \frac{T - bN_a}{\sqrt{T + N_{a}^2 * \sigma_b^2/N_{sky}}}
\end{equation}
where $T$ is the total number of counts inside the aperture, $b$ is estimated background
counts per pixel, $\sigma_b$ is the measured dispersion in $b$, 
in units of counts pixel$^{-1}$, $N_a$ is the number of pixels inside the
aperture, $N_{sky}$ is the total number of sky pixels used to estimate the background. The 
quantity $\sigma_b^2/N_{sky}$ is the square of the uncertainty in the mean background counts. 
Rather than assume the background was Poisson, we measured the dispersion directly. (The OM
background is close to Poissonian based on this comparison.)
A $3\sigma$ upper limit was computed for each aperture based on the Poisson
uncertainty and the uncertainty in the mean background level, and in cases where the net 
counts did not exceed this background, we report this $3\sigma$ upper limit. 

Photometric errors, including Poisson uncertainties and background subtraction, dominate.
Flat fielding errors are estimated to be a few percent. There is also a systematic 
absolute calibration uncertainty of $\sim3\%$\footnote{Talavera, A.,  OM Cal Team 
XMM-Newton Optical and UV Monitor (OM)
Calibration Status, XMM-SOC-CAL-TN-0019, Issue 5.0, 30 Oct 2008. \url{xmm2.esac.esa.int/docs/documents/CAL-TN-0019.ps.gz}}. 
To account for these systematic errors in estimating colors based on photometry
from other observatories, we add a magnitude uncertainty of 0.05
in quadrature to all UV-optical colors in our analysis and plots. Errors reported for photometry in  
Table~\ref{table:omphot} are statistical only. 


We utilized cross-matches of our BCG locations with the source catalog from the
Galaxy Evolution Explorer (GALEX) data archive General Release 4 (GR4).\footnote{The GALEX archive
is hosted by the Multi-Mission archive at the 
Space Telescope Science Institute (MAST).} The near ultraviolet filter (NUV, 231 nm) has a similar center to the UVM2 OM
filter, but is broader, and the far ultraviolet filter (FUV, 153 nm) is centered at a shorter wavelength 
(Figure~\ref{templates}). The GALEX
telescope is more sensitive than the OM (in the UV), but typical GALEX exposures are far shorter and
do not cover our full sample. The GALEX PSF ($\sim4-5\arcsec$) is significantly poorer than
the OM's PSF. Seventeen of the 25 BCGs with GALEX observations were matched in the
GALEX catalog to better than $6\arcsec$, (11 of these with magnitude uncertainties better than 0.3 mag). 
We inspected each GALEX field, including images lacking BCG detections. Low-confidence sources from the GR4 catalog
were rejected based on this inspection, and their fluxes were converted to upper limits. Two
low-confidence sources remain on the detection list (RXCJ0821.8+0112 and RXCJ2218.6-3853) because of possible extended, but
low surface brightness, emission visible in the image. These two sources are flagged in Table~\ref{table:galexphot}. 
The remaining sources were either detected robustly 
(magnitude uncertainties $\leq 0.3$ mag) or were detected in both bands. We do not base any strong
conclusions on the GALEX data or the UVM2 data because of the large number of upper limits.

For at least one NUV/FUV 
detection (the BCG in RXCJ2234.5-3744), the GALEX magnitude represents a blended source and is too bright.
This source is flagged in the tables and plots where it appears.
The GALEX catalog contains aperture magnitudes for radii from $1.5\arcsec-18\arcsec$. We interpolate these magnitudes and 
uncertainties to match the OM and SOAR-R apertures of $10 h^{-1}$ kpc. 
Six of the eleven had FUV detections with better than 0.3 magnitude
uncertainties. No believable FUV detection was found for sources lacking NUV detections. RXCJ2014.8-2430 lacked a 
corresponding FUV observation. We report these aperture measurements and total flux estimates 
in Table~\ref{table:galexphot}. The upper limits of sources observed 
by GALEX are $\sim m_{NUV} < 21$ for the shallow, wide-area all-sky survey (AIS) with exposures of
$\sim100-200$ seconds, not too different from the OM upper limits; however four of our GALEX fields have much
deeper observations.

\section{Optical Spectroscopy}
\subsection{SOAR Goodman Spectrograph}

Optical spectra of the objects were obtained with the newly installed 
Goodman Spectrograph \citep{2004SPIE.5492..331C} on the 4.1m Southern Astrophysical Research (SOAR) Telescope in Cerro Pach\'{o}n, Chile.  The observations were taken between September 2008 and May 2009.  For all objects, a $1.68\arcsec$ 
wide slit and a 600 lines-per-mm grating was used, for a wavelength coverage of 640-915 nm.  Beginning 2009 April 17, a blue blocking filter (GG-495) was used to prevent second-order scattering of blue light onto the CCD, which was not a serious issue for any of these data.  All data were taken during photometric observing conditions. All observations were taken under excellent conditions, nearly photometric, and good seeing ($<1"$). 
Our strategy was to observe each object twice: once along the major axis and once along the minor axis.  Due to hardware limitations for rotating the spectrograph, a fully perpendicular observation was not always possible.   Information about the observations can be found in Table~\ref{table:goodman}.  Observations of spectrophotometric standard stars \citep{1992PASP..104..533H} were made 
for flux calibration (Table~\ref{table:goodman}).

Data reduction consisted of CCD overscan subtraction, trimming the image. 
The Goodman spectrograph suffers from $20\%$ peak-to-peak 
spectroscopic fringing redward of about 7000 \AA. We used normalized quartz flats 
taken directly after the completion of each exposure to remove
this fringing signature. However, the fringing pattern shifts and distorts, 
probably due to flexure of the instrument. The fringing redward of 8000 \AA~ was
the most difficult to remove. Since the wavelength scale of a fringe 
feature was broader than a typical emission line, it was straightforward to determine which sources had emission lines
brighter than $\sim1-2$ \AA~ in equivalent width. Wavelength solutions were obtained
based on night sky lines. We measured redshifts, line widths, and equivalent widths, and estimated
the H$\alpha$ line luminosity.

We extracted single spectra with widths of 20-40 pixels, corresponding to $3-6\arcsec$ in the spatial direction.
We detected emission lines (H$\alpha$, [NII]6584/6548\AA, [SII]6717/6730\AA, and usually [OI]6300\AA) from 7 BCGs out of 31. (The
32nd, the BCG in RXCJ1311.4-0121, also known as Abell 1689, is discussed in the next section.)  
We report upper limits and H$\alpha$ equivalent width detections
in Table~\ref{table:goodman} and emission line properties in Table~\ref{table:emissionlines}.
Error estimates incorporate counting statistics and fit uncertainties based on bootstrap simulations
of the spectra, accomplished with the task {\it splot} in IRAF \citep{1993ASPC...52..173T}. 
The residual fringing modulates red Goodman spectra continua on wavelength scales somewhat
broader than typical linewidths, and it represents the most serious source of systematic uncertainty. 
Upper limits are thus typically 1-2 \AA. For H$\alpha$ emission lines, underlying stellar absorption can affect
the total flux estimate. We have added 1 \AA~ of uncertainty in quadrature to our H$\alpha$ equivalent width uncertainties.
This uncertainty only matters for the faintest systems.
The two  most prominent emission line systems, RXCJ1141.4-1216 and RXCJ2014.8-2430, were also the bluest in 
$UVW1-R$ and were the two best detections in the bluest OM UV filter, UVM2.   
RXCJ2014.8-2430 and RXCJ1141.4-1216 were also the brightest 
GALEX NUV sources, $19.84\pm0.085$ and $19.68\pm0.09$ in AB magnitudes. They represent
the most luminous GALEX NUV sources in REXCESS as well.

\subsection{Las Campanas DuPont Telescope Modular Spectrograph}

To complete the spectroscopic coverage of a relevant comparison sample of X-ray bright clusters of galaxies, the
Brightest 55 X-ray Cluster sample (B55; Piccinotti et al. 1982),  we
report here upper limits for BCG H$\alpha$ emission lines, as obtained using the Modular 
Spectrograph\footnote{\url{http://www.lco.cl/lco/telescopes-information/irenee-du-pont/instruments/website/modspec-manuals/modular-spectrograph}}
 on the 2.5 meter DuPont Telescope on Las Campanas. These observations were taken and reduced as described in
\citet{1993ApJ...414L..17D}. One of these clusters, Abell 1689 (RXCJ1311.4-0120), is also in the 
REXCESS sample. We provide here upper limits of H$\alpha$ equivalent widths of 
$<1$\AA~ for four additional BCGs in clusters with 
unknown emission-line properties: Abell 1650, Abell 3571, Triangulum Australis, and Abell 4038.
Abell 4038 is also known as Klemola 44. Our results are reported in Table~\ref{table:lco}.
Other optical emission line measurements for the B55 sample are provided in Peres et al. (1998), 
Crawford et al. (1999), and Cavagnolo et al. (2009). Non-detections of emission-line signatures from 
3C129 and Abell 3532 (Klemola 22) are reported in  \citet{1975ApJ...199L...1S} and \citet{1987A&A...179..108C}, respectively.

\section{Ultraviolet - Optical Colors and UV Excess Estimates \label{section:typ} }

\begin{figure}
\plotone{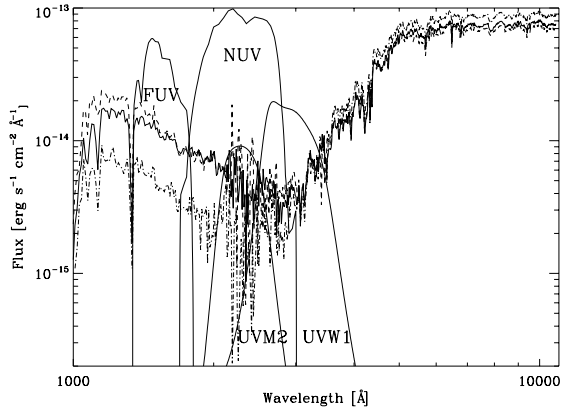}
\caption[]{Filter throughput profiles, used for the UV photometry reported in this paper, plotted against
the redshifted ($z=0.1$) spectra of three elliptical galaxies, M49, M60, and NGC1399, provided by Thomas Brown. Note
that the GALEX NUV filter and the OM UVM2 filter have similar mean wavelengths, but different widths. The relative heights
of the two sets of filters (OM vs. GALEX) are qualitative; the GALEX system is significantly more sensitive to UV than the
Optical Monitor system.   \label{templates}}
\end{figure}

\begin{figure}
\plotone{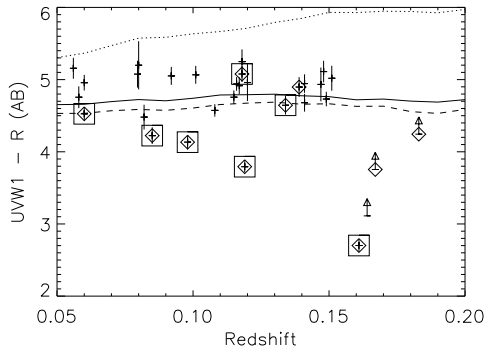}
\caption[]{UVW1-R AB colors and 3$\sigma$ lower limits. Overplotted are the colors predicted from template
spectra of three nearby elliptical galaxies with a range of UV colors, whose spectra are shown in
Figure~\ref{templates}. The dotted line is for M49.  The 
solid line is for M60, and dashed line is for the colors of NGC1399. Diamonds are overplotted on points representing BCGs in 
cool-core clusters, as identified by Pratt et al. (2009). 
The boxes identify galaxies with detected H$\alpha$ emission from this work.
\label{UVW1}}
\end{figure}

\begin{figure}
\plotone{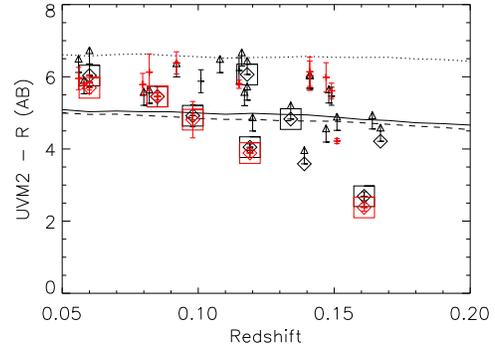}
\caption[]{Observed 
$NUV-R$ AB colors are plotted in red and $UVM2-R$ AB colors and 3$\sigma$ lower limits on UVM2-R colors are plotted in black, as
a function of redshift.   Overplotted are the colors predicted from template
spectra of three elliptical galaxies with different UV colors, same line codes and point conventions as for Figure~\ref{UVW1}. 
The correction between the GALEX NUV filter and the XMM OM UVM2 filter is small, $\pm0.1$ magnitudes for the template galaxies. However
the NUV filter is significantly broader than the OM UVM2 filter.
The blue NUV-R outlier with tiny error bars at $z=0.15$ is for the BCG in cluster RXCJ2234.5-3744, which in GALEX is a blend of 2-3 UV sources, the
brightest of which is unlikely to be associated with the BCG. The
OM upper limit, directly above it in the graph, is for the BCG alone.
\label{UVM2}}
\end{figure}

\begin{figure}
\plotone{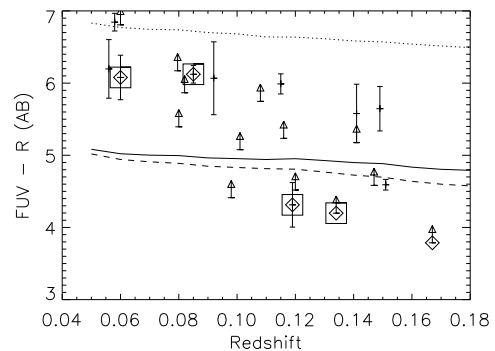}
\caption[]{Observed FUV-R AB colors plotted versus redshift for the 11 BCGs with detections and 13 BCGs with 
lower limits in GALEX FUV observations.  Overplotted are the colors predicted from template
spectra of three nearby elliptical galaxies with a range of UV colors, with spectra shown in
Figure~\ref{templates}. Points and lines are coded as in Figure~\ref{UVW1}. The point at $z=0.15$ with tiny error bars is that of 
a blended GALEX source (RXCJ2234.5-3744), and is not likely to be the color representative of the BCG.
\label{FUV}}
\end{figure}

Examples of UV spectra of low redshift ellipticals, M60, NGC1399, and M49, 
kindly provided by Thomas Brown, are shown in Figure~\ref{templates}, along with the wavelength
coverage of the four UV filters used in this work, OM-UVM2, OM-UVW1, GALEX-NUV, and GALEX-FUV.
The observed UVW1-R, UVM2-R, and FUV-R colors are plotted as a function of redshift in
Figures~\ref{UVW1}, ~\ref{UVM2}, and \ref{FUV}. Figure~\ref{UVW1} includes
NUV-R colors for BCGs detected with GALEX, plotted in red. 

The measured UV-optical colors of the REXCESS BCGs exhibit a scatter beyond that expected
from measurement uncertainties alone. We assess a systematic uncertainty in the UV 
zero point magnitudes of $0.05$ magnitudes, added in quadrature to 
the statistical uncertainties reported in Table~\ref{table:omphot}.
The average UVW1-R AB color for the galaxies without H$\alpha$ emission  
and with both R and UVW1 measurements (21 galaxies) 
is $4.69\pm 0.33$. This mean excludes the two bluest BCGs in UV-optical 
colors ($UVW1-R < 4$, RXCJ2014.8-2430 and RXCJ1141.4-1216), and
the other 5 BCGs with emission lines. 
Even with these exclusions, the
scatter significantly exceeds the measurement error. 
For all 28 galaxies with R and UVW1 measurements, the average UVW1-R color is $4.54\pm 0.56$. 
The colors of the non-emission-line BCG sample are consistent with the redshifted template spectra
of nearby ellipticals. Only the colors of galaxies with emission lines, and
presumably recent star formation, deviate significantly from the colors of the templates (Figure~\ref{UVW1}). 
The same conclusion is true for the small sample of our clusters with bluer NUV and UVM2 colors (Figure~\ref{UVM2}).

One might worry that emission lines may contaminate the R-band measurement. We note that the two 
BCGs with the most excess UV ($UVW1-R < 4$) and brightest H$\alpha$ emission 
are among the brightest BCGs in the sample, RXCJ1141.4-1216 and RXCJ2014.8-2430, with
rest-frame R magnitudes of -23.4 and -23.2, respectively. But the least luminous BCG
in our sample, RXCJ2319.6-7313 with a rest-frame R magnitude of -21.8, is also an emission-line system.
The summed equivalent widths of the emission lines (only a few of which 
fall into the R bandpass) are quite small compared to the bandwidth
of the R filter.

It is illustrative to consider the huge impact on the UVW1-R color of a tiny
amount of star formation, predicted by stellar synthesis models. 
For example, Han et al. (2007) derive spectral energy distributions (SEDs) for
a single age population wherein mass transfer in binary stars is the mechanism
for producing the UV-upturn stars in the old population. The color of the
15 billion year old population is $UVW1-R \sim 5.6$ ($UVM2-R \sim 6.8$), compared to a 100
million year old population color of $UVW1-R \sim 1.2$ ($UVM2-R \sim 1.3$). 
A very small population of newly formed stars has a dramatic impact on the
UV-optical colors. 

Recent UV results for M87 from the Advanced Camera for Surveys Solar Blind Channel 
suggests that some of the far UV
could even come from [C~IV] (1549~\AA) emission \citep{2009ApJ...704L..20S}. The GALEX FUV fluxes could also
be affected by the presence of Lyman $\alpha$, for the $z>0.15$ sources; [C~IV] may contribute
to flux in the broad NUV filter. However, as we will show in \S6, 
the lack of a redshift trend in colors suggests contributions from UV emission lines 
are insignificant, at least for this sample of BCGs.

\section{Discussion}

For the purposes of the following discussion, we classify galaxies as active if they host an
emission line system with  H$\alpha$ equivalent width $ \gtrsim 1$ ~\AA, and
as inactive if they don't.

\subsection{UV Emission from Brightest Cluster Galaxies}

Because we have reliable X-ray measurements for every cluster and because the
X-ray sample is selected specifically to calibrate scaling relationships, 
our sample is particularly well-suited  
for studies of possible relationships between BCG star formation and BCG UV-optical colors, 
but also among BCG mass, luminosity, cluster mass, and cluster dynamic state. Our 
emission-line based classification of inactive vs. active 
is consistent with that used to classify elliptical galaxies for GALEX galaxy studies. 
Although our original intent for our BCG study was to examine the stellar mass \citep{2010ApJ...713...1037H} 
and star formation signatures (this paper) 
in these BCGs, the scope of this paper expanded to include quantifying the contribution of the
UV from the older population, since an excess UV attributed to star formation must be over and above
that contributed by old stars. That we detected nearly all of the BCGs in the UVW1 band of the XMM OM 
means at least some of them are dominated by old stellar populations.
 
All BCGs seem to have at least some UV emission despite their ``red and dead'' reputation.
Elliptical galaxies exhibit differing amounts of UV shortward of 2500\AA, sometimes called the UV upturn, because
$L_\lambda$ slopes upward to shorter wavelengths (see \citet{1999ARA&A..37..603O} for a review and references.) 
\citet{2000ApJ...532..308B, 2002ApJ...568L..19B}, using HST observations, showed
the culprits for this emission turn out to be extreme horizontal branch stars (eHBs).   
The UV continuum light from galaxies therefore can be produced by two different populations of stars:  
(1) the most massive, recently formed stars and (2) eHBs. In order
to assess the amount of excess UV attributable to recent star formation, we need to establish the baseline contribution
of the old population. This task is not as easy as it might sound.  
As long as these old UV stars are numerous, the UV-optical colors of 
an inactive BCG should be relatively constant for BCGs of similar 
metallicity, but observations show that the UV colors of ellipticals exhibit scatter. One explanation is
that the UV-optical color of an elliptical galaxy  
is exquisitely sensitive to any star formation in the last 100 billion years. Studies of inactive
elliptical galaxies show that tiny amounts of star formation can induce scatter in UV-optical or near
infrared colors as large as observed, even though their line emission is below detection thresholds 
similar to those of our study \citep[e.g.][]{2005ApJ...619L.111Y,2007ApJS..173..619K,2008MNRAS.385.2097R}. 
The REXCESS
sample includes BCGs with and without line emission, so we discuss the implications of our data
for both populations.

Commonly-used population synthesis codes, such as \citet{1993ApJ...405..538B} or
Starburst99\footnote{\url{http://www.stsci.edu/science/starburst99/}} \citep{1999ApJS..123....3L}, 
do not predict  
the UV spectrum of an old stellar population \citep[e.g.,][]{1993ApJ...417..102M}. Therefore, in 
order to explore the baseline UV contributions of an old population, we 
turned to empirical UV spectral templates (M60, M49, and NGC~1399) to generate the inferred observed 
broadband colors as a function of redshift in Figures~\ref{UVW1}, \ref{UVM2}, and \ref{FUV}. 
{\em Our results suggest that BCGs hosting old populations with 
little to no star formation have a limited range of UV-optical colors, 
and that BCGs with UV-optical colors bluer than this range host  
the strongest emission-line systems, and presumably, recent star formation.} Whether
the latter is true or the H$\alpha$ emission comes from a low-luminosity AGN, accretion of
cold gas by the BCG has taken place. 

The elliptical galaxies for which we have templates 
host old stellar populations, as confirmed by the presence of H$\beta$ equivalent widths,
1.4--1.6 \AA~ in absorption \citep{2000MNRAS.315..184K,2001MNRAS.323..615K}.
The templates demonstrate that over the redshift range of our sample, there is very little color change
arising from shifting the baseline spectrum as a function of wavelength, 
so we do not k-correct our UV magnitudes, in order to stay as close to the data as possible. 
The templates also show that for a $z=0.1$ galaxy, the 
UVW1 bandpass sits at the local minimum between the UV upturn feature shortward of 250 nm
and the near-UV (250 nm rest) emission from normal stars (Figure 1). Therefore the sensitivity of the UVW1-optical 
color to the behavior of  eHB stars is minimized compared to UV-optical colors based on the GALEX filters. 
The templates also show variation in this color between them.

The inactive BCGs in our sample seem to represent a sample of galaxies with 
less dispersion in their UV-optical colors than that of ellipticals as a whole, and 
their colors are consistent with colors of inactive ellipticals. Whether this uniformity means
that they have much less contamination from star formation, or are more uniform in their 
metallicity and age than their lower-mass cousins, is an open question. 

On the other hand, the BCGs in our sample {\em with} emission-lines show a consistent correlation between  
excess UV emission (over that expected from an old population) 
and the strength of the H$\alpha$ emission line. This correlation  
suggests that star formation is occurring in
the subset of BCGs where we expect star formation \citep[e.g.,][]{2008ApJ...683L.107C,2008MNRAS.385..757D}: 
the cool cores with the shortest central ICM
cooling times. The UV-optical colors of these objects therefore appear to be sensitive to star formation.

\subsection{Correlation of H$\alpha$ Equivalent Width with Cluster Properties and UV-Optical BCG colors \label{emcol}}

\begin{figure}
\plotone{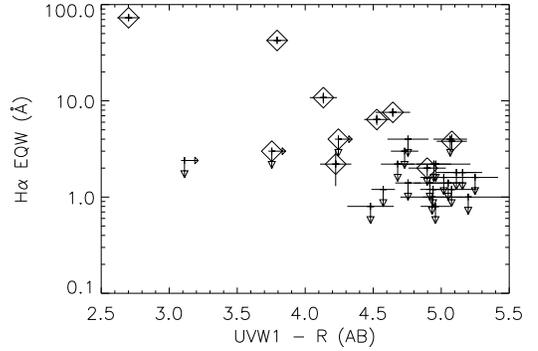}
\caption[]{UV-R AB colors vs. H$\alpha$ equivalent width.
Diamonds are overplotted on points representing BCGs in 
cool-core clusters, as identified by Pratt et al. (2009).  \label{HaUVOpt}}
\end{figure}

BCGs with a UV excess over and above what is expected from an old population are more likely to exhibit
H$\alpha$ emission. Furthermore, the line strength of H$\alpha$ is strongly correlated with the UV excess 
(Figure~\ref{HaUVOpt}). 

Seven BCGs in our sample exhibited detectable H$\alpha$ emission (also [N~II]6548\AA, 6584\AA, 
[S~II]6717\AA, 6731\AA, and occasionally [O~I]6300\AA).
All of these BCGs inhabit clusters identified as cool cores (CC) in Pratt et al. (2009), and all of these
are associated with a 1.4 GHz radio source. All 3 CCs in Pratt et al. (2009) that
were not detected in H$\alpha$ have central cooling times estimated from XMM observations (at $0.03R_{500}$) 
longer than $10^9$ years, while only 2 out of 7 in the detected sample have cooling times this long.

Six of these seven emission-line BCGs also showed a UV excess, as defined by having a UVW1-R color 
significantly bluer (UVW1-R$ < 4$ or NUV-R$ < 5.7$) than that expected from ellipticals with
no star formation (UVW1-R$\sim 5$, NUV-R$ \sim 6$). The brightest two of these were the only two galaxies in the 
sample with detected OM UVM2 flux  at a statistical confidence $>5\sigma$. Five of
the six emission-line galaxies with public GALEX NUV observations were detected in the NUV,
whereas 11 of the 17 non-emission line BCGs with such observations were detected, and none
of them with $m_{NUV} < 20$.

\begin{figure}
\plotone{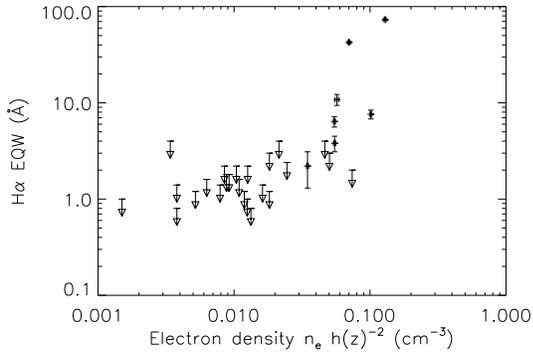}
\caption[]{Scaled central electron density ($h(z)^2$ cm$^{-3}$) \citep{2008A&A...487..431C,2010ApJ...713...1037H}
 vs. H$\alpha$ equivalent
width in Angstr{\"o}ms. The probability that these two quantities are not correlated is $<0.0001$, according to various survival 
statistics utilizing upper limit information. 
\label{nelecha}}
\end{figure}

\begin{figure}
\plotone{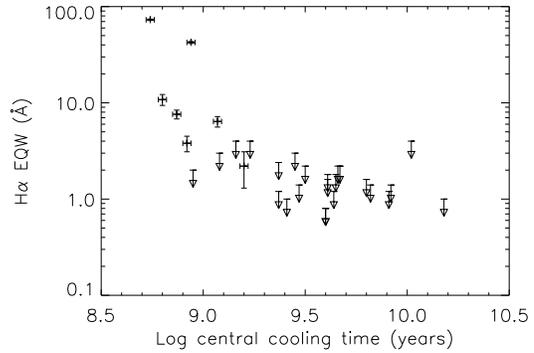}
\caption[]{Log of central cooling time \citep{2008A&A...487..431C,2010ApJ...713...1037H} in years vs. H$\alpha$ equivalent
width in Angstr{\"o}ms. The probability that these two quantities are not correlated is $<0.0001$, according to various survival 
statistics utilizing upper limit information. 
\label{tcoolha}}
\end{figure}

\begin{figure}
\plotone{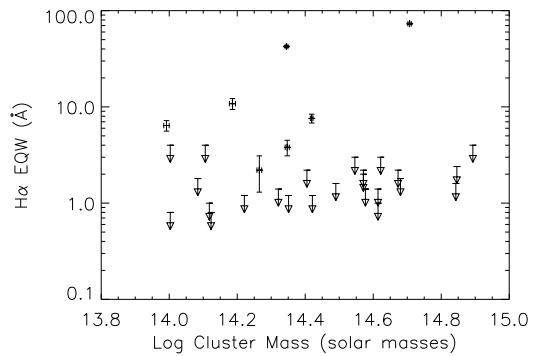}
\caption[]{Log of gravitating mass in solar masses, based on X-ray determinations \citep{2010A&A...511A..85P} vs. H$\alpha$
equivalent width in Angstroms. These two quantities are not correlated. For comparison with previous correlation
probability estimates, the probability that these two quantities are not correlated is $\sim0.33$. 
\label{massha}}
\end{figure}

Cluster ICM core properties of electron density ($n_e h(z)^{-2}$) and cooling
time are strongly correlated with H$\alpha$ equivalent widths.
In Figures~\ref{nelecha}-\ref{massha} we show two strong correlations (with central
electron density and cooling time), and, for
comparison, a weak correlation (with cluster mass). We explored several tests
of correlation, using tasks in the survival statistics package in IRAF/STSDAS 
\citep{1986ApJ...306..490I}.
Survival statistics utilize information in the upper limits as well as the
detections. The measures of correlation (or lack thereof) were consistent from test to test. 
A generalized Kendall's $\tau$ test and Spearman's $\rho$ test 
showed the probability that the equivalent width of H$\alpha$ was
not correlated with central electron density or cooling time was $<0.0001$. A similar
set of tests showed that the probability the equivalent width of H$\alpha$ was {\em not}
correlated with the UV-optical color was $1-2\%$.  
Similarly, pairing UV-optical colors with central electron density or cooling time produced
probabilities of $\sim0.02$ that the paired properties were not correlated. 
The UV correlations with H$\alpha$, central electron density, and cooling time 
are undoubtedly diluted by the fact that most of the galaxies in
our sample have UV emission dominated by an old stellar population.
For contrast, the same statistics for the correlation of X-ray mass and H$\alpha$ equivalent 
width gave a probability $\sim0.33-0.38$ that it was not correlated, which confirms
the visual impression.

\begin{figure}
\plotone{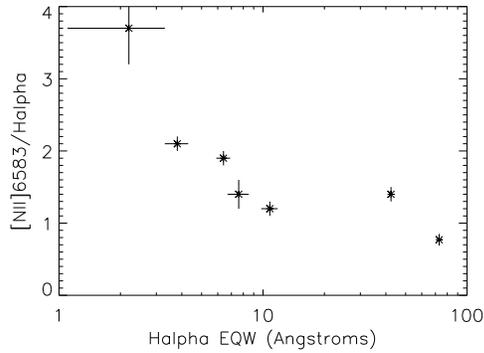}
\caption[]{[NII]6583/H$\alpha$ ratio of integrated line fluxes plotted against H$\alpha$ equivalent width in Angstroms. 
The trend appears to be that more luminous emission line systems have [NII]/H$\alpha$ ratios $\sim1$, which increase
as the emission becomes 
less luminous. The faintest emission-line system plotted here may be affected by intrinsic (stellar) H$\alpha$ 
absorption. The [NII]/H$\alpha$ ratio is determined by the temperature of the emission line gas relative to the
recombination rate. If the gas is hotter than can be achieved by photoelectric heating by stars, typical
photoionization models can generate [NII]/H$\alpha > 1$ \citep[e.g., ][]{1991ApJ...381..361D}. 
It is interesting that the H$\alpha$ systems with the lowest equivalent widths exhibit the
highest [NII]/H$\alpha$ ratios, which in turn are found in systems with typical UV-optical colors. This pattern
suggests that the lowest emission-line luminosity systems may not be dominated by star formation-related processes. \label{NIIHalpha}}
\end{figure}

We computed the [NII]6583/H$\alpha$ ratios for the seven BCGs with H$\alpha$ emission. We note that the [NII]/H$\alpha$ ratio is unusually high
for HII regions, but in the range observed for cooling flow BCGs \citep[e.g.][]{1989ApJ...338...48H} (Figure~\ref{NIIHalpha}). This ratio, which
is sensitive to the temperature and excitation of the emission line gas, trends with the relative strength of the H$\alpha$
line, in equivalent width here. The most luminous H$\alpha$ systems have lower [NII]/H$\alpha$ ratios than the weak
systems. The less luminous emission line systems may be increasingly dominated by excitation processes such as cosmic ray
heating, conduction, and even weak shocks \citep[e.g., ][]{2008MNRAS.386L..72F}, while the more luminous systems may have proportionally more heating from
stars. Forbidden line emission can be elevated in photoionized systems with additional sources of 
heat \citep[e.g.][]{1997ApJ...486..242V}. This interpretation is consistent with the 
correlation of UV-optical color and H$\alpha$ equivalent width. For this work, we interpret H$\alpha$ and
forbidden line emission as signs of activity, at least some of which is associated with recent star formation,
but we note the many alternate sources of H$\alpha$ that may be present in these systems
\citep{2008MNRAS.386L..72F, 1989ApJ...338...48H, 1990ApJ...360L..15V, 1996MNRAS.280..438J,1990MNRAS.244P..26B}.

\subsection{Incidence and Rate of Star Formation in Cluster Samples}

The star formation rates based on a UV excess or an H$\alpha$ luminosity in terms of a specific star formation
rate are model-dependent. H$\alpha$ emission is sensitive to the most recent star formation, since
only the hottest, shortest-lived stars ionize hydrogen. H$\alpha$ rates are therefore closer to
being an instantaneous measure of star formation. In contrast, the UV is sensitive to a broader
range of hot stars, and therefore the age of the burst affects the magnitude of the star formation
rate. The UV is easily obscured by gas and dust, and represents a lower limit, modulo the age assumption. Furthermore,
the lack of certainty about the exact level of UV light contributed by eHBs affects the 
estimate of UV excess for the fainter systems (the level of the baseline is irrelevant for strong UV excesses.)
Finally, any given amount of UV excess light can be associated with a nearly limitless amount
of stars associated with a single burst, since a tiny amount of UV can represent the UV tail of a huge burst,
say $10^8$ years ago. The UV rates computed here are based on an assumption of continuous
star formation, which provides SFRs at the low end of the scale \citep{1998ARA&A..36..189K}.

The excess UV of the two bluest galaxies in UV-optical light 
is consistent with that emitted by an {\em unobscured} population continuously forming stars for timescales
longer than $10^8$ years, at about 2 solar masses per year for RXCJ1141.4-1216 and
8 solar masses per year for RXCJ2014.8-2430, based on a UV excess over a baseline color of $UVW1-R=4.7-5$ 
\citep{1998ARA&A..36..189K}. Similar star formation rates are inferred from relations produced by 
models in \citet{2003A&A...410...83H}, in which the onset of constant 
star formation ranges between $10^7 - 10^8$ years in the past 
(Table~\ref{table:emissionlines}). The rates for the more recent times of onset are
about 60\% higher because the UV luminosity increases to a maximum at a population
age of $10^8$ years \citep{2003A&A...410...83H}. These rates are underestimates, since
we have not taken into account attenuation of the UV by the interstellar medium of the host galaxy.

H$\alpha$ is produced by ionization by the hottest, youngest stars, and therefore 
the star formation rate based on H$\alpha$ emissions is an indicator of the nearly
instantaneous star formation rate. Table~\ref{table:emissionlines} shows that the H$\alpha$ rates
are consistent with the range of derived UV rates. The approximate diameter of the
brightest two emission lines systems is $\sim10$ kpc (of order twice the slit width), 
similar to that of the emission line nebula in Abell 2597
\citep{1997ApJ...486..242V}. Given the unknowns in the estimation process 
(attenuation, star formation history, slit for an extended source H$\alpha$ measurement), any
differences in the rates can be easily explained. 
For example, somewhat higher UV rates can be explained by an older starburst and
slit losses, and somewhat higher H$\alpha$ rates can be explained by moderate UV attenuation.
Therefore, to first order, the production of UV (in excess of that from an old population) 
and optical emission lines in 
these systems is not pathologically different from that seen in star-forming galaxies. 

The main result is that the appearance of emission lines together with a UV excess is consistent with
star formation being the source of both. None of these galaxies has an X-ray or optically-bright AGN; 
however, since 8 out of 10 (9 out of 11 if one includes RXCJ0211.4-4017 as a cool core) of the
cool core clusters in this sample have radio loud AGN, some of the H$\alpha$ may 
be associated with the AGN. For example, cosmic-ray heating from relativistic particles 
may be important in generating some of the H$\alpha$ emission \citep{2008MNRAS.386L..72F}, especially in the weaker systems, 
in which the larger [NII]/H$\alpha$ ($>1$) ratios suggest that the interstellar gas may be 
hotter than  typical of regions photoionized by stars. 
We showed in the previous section that the H$\alpha$ equivalent
width is correlated with the UV excess; so at least some of the H$\alpha$ luminosity is likely generated
by UV-producing stars. 

The incidence of emission-line sources in the REXCESS sample (22\%)
is slightly low compared to the incidence of H$\alpha$-emitting
systems in other low-redshift cluster samples, but not remarkably so. 
The B55 sample \citep{1982ApJ...253..485P,1998MNRAS.298..416P}, 
including the spectroscopic observations reported in Table~\ref{table:lco}, 
has an incidence rate of $38\%$ for the now complete sample. 
For the sample of X-ray clusters in \citet{1999MNRAS.306..857C},
the incidence was $31.5\pm4\%$. \citet{2005MNRAS.359.1481B} show that in the BCS/EBCS 
\citep{1998MNRAS.301..881E} sample with $z>0.15$, 13 of the 33 clusters with public {\em Chandra}
data and available optical spectroscopy had H$\alpha$, an incidence rate of $39\%$.
A study of luminous
Extended Medium Sensitivity (EMSS) X-ray clusters \citep{1991ApJS...76..813S,1990ApJS...72..567G}, 
at $z=0.07-0.4$ from the Einstein Observatory,  
\citep{1992ApJ...385...49D}, showed rates
of 40-50\% in a flux-limited, complete, but high X-ray luminosity, sample. They  
reported only weak evidence for lower incidence at the highest redshifts of this study, 
so there was very little evidence for evolution in the incidence rate for $z<0.4$. 
In contrast, \citet{2007hvcg.conf...48V} claim there is a
lack of X-ray evidence for cool cores at high redshift ($z>0.5$), which implies also that
very few high-redshift BCGs will host emission line systems.

BCGs of optically-selected clusters have a much lower incidence rate of emission lines.
\citet{2007MNRAS.379..867V} found  an incidence rate of $\sim15\%$ in a study of 625 Sloan Digital Sky Survey BCGs in optically-selected
clusters \citep{2005AJ....130..968M}  with a control sample
matched in absolute magnitude, redshift, and color. This rate was  
not much higher than that of the control sample. However, this study omitted the most massive BCGs
because of the lack of truly massive galaxies in their control sample. \citet{2007MNRAS.379..100E} examine
93 clusters (selected somewhat heterogeneously, including both X-ray selected clusters and
optically selected C4/DR3 clusters) between $z=0.010-0.067$. Of these, $15\%$ had emission lines, with no trend in
galaxy magnitude or velocity dispersion. Of the ``cooling flow" clusters, 
$71\pm^{9}_{14}\%$ had optical emission; if those BCGs were within
50 kpc of the X-ray centroid, 85-100\% had optical emission. If the cooling flow
clusters were omitted, only 10\% of the sample had any line emission, similar to the control sample.

Statistically, the incidence rate in the REXCESS sample is somewhat lower than that of other 
X-ray samples (B55 and EMSS, for example), and similar to that of optically selected samples.
The binomial probability of finding seven or fewer BCGs with emission lines given an
expectation of 15\% is about 9.8\%, statistically
consistent with the incidence rate in optically-selected cluster surveys. If the expectation of detecting emission
lines is 35\%, the probability of finding seven or fewer is 4.5\%, indicating 
a result about $2\sigma$ different from
expectations for X-ray selected surveys. REXCESS therefore has a somewhat lower number of BCGs with optical emission
lines than the B55 \citep{1999MNRAS.306..857C} or EMSS \citep{1992ApJ...385...49D} samples. The
incidence of emission line systems inside the cool-core sample (7 out of 10) is certainly consistent with the
high incidence ($\sim70\%$) found in \citet{2007MNRAS.379..100E}.

We note that the comparisons presented here are not of cluster samples matched in mass.
The number density of clusters in the wide-area optical surveys is high, 
so their average masses must be lower and the range of masses sampled greater
 than the X-ray selected surveys under discussion here. The B55 and EMSS samples are complete,
 flux-limited samples, which REXCESS is not. We do not see, to the limit of our small detection
 statistics, evidence for some dependence of the incidence rate of emission-line BCGs 
 on the mass of the cluster, but the REXCESS sample does not include low mass clusters or groups.

The explanation - if one is needed for a 2$\sigma$ discrepancy  - is not clear. X-ray flux-limited samples
have on average higher-mass clusters than the larger optically-selected samples. They 
may preferentially include objects with higher gas mass fractions for a given mass or objects with
X-ray bright cool cores \citep{2009A&A...498..361P,1994MNRAS.267..779F,1992MNRAS.258..177E}. 
Analysis of the REXCESS sample itself shows 
no evidence whatsoever of any mass dependence of the presence of a cool core
(B{\"o}hringer et al., in preparation). Certainly, the EMSS may have included a few more
cool-core clusters because of its fixed-size sliding-box selection method, as discussed in \citet{1992ApJ...385...49D}.
However, the B55 survey was selected by collimator flux, and therefore is unlikely to be biased towards
cool core clusters. The REXCESS selection had no specific morphological criterion, so one might
expect the selection to be similar to that of the B55 sample. The relatively small sample size, however, 
does not allow hard conclusions to be drawn, since the similarity in REXCESS BCG emission-line incidence to
that of an optically selected sample could be coincidental.
Other X-ray samples are limited in their size, dynamic
range in mass or redshift, and the degree to which the sample contents are representative of
the full population of clusters of galaxies \citep{2007hvcg.conf...48V,2008A&A...483...35S, 2007ApJ...657..197S}. 
 
 The discrepancy in the emission-line BCG incidence rates between optically- and X-ray-selected 
 samples does not appear to have a ready explanation in terms of incidence rates varying with
 cluster masses, X-ray temperatures, or BCG optical luminosity, at least to the limits probed by the REXCESS 
 cluster sample. However, we do see a strong correlation with the presence of low entropy gas, short
 central ICM cooling times, high central electron densities, and, as we will show below, relaxed
 dynamical state. If optical samples include a larger proportion of unrelaxed, disturbed clusters and
 a smaller proportion of clusters with cool cores, then the lower incidence rates of emission-line BCGs
 in optical samples is easier to understand. Larger and more inclusive studies,
with attention to sample selection, are required for further progress in exploring the incidence
of star formation in BCGs of clusters of galaxies of all types, in order to understand how and why  
the thermodynamic state of the intracluster gas is coupled to star formation in the BCG.

\subsection{Are Inactive BCGs Nearly Identical?}

For reasons unknown, BCGs are surprisingly good 
standard candles, at least at low redshifts \citep{1995ApJ...440...28P}, 
and inactive BCGs may be an even more homogeneous set.
In the following sections we will explore the apparent similarity of the UV and optical
properties in the inactive BCGs in our sample, as well as the tendency for 
outliers to host H$\alpha$ emission line systems.  We will show that the UVW1-R color,
a quantity sensitive to metallicity and age of the stellar populations, is 
 insensitive to BCG R-band luminosity and color in a metric aperture and to the redshift
 range spanned by our sample. Because the REXCESS clusters  have homogeneous
and high quality X-ray measurements, we are also able to show that the UVW1-optical colors of inactive
BCGs do not vary with halo gas temperature (equivalently cluster halo mass) or dynamical state. 

\subsubsection{Lack of Trends in UV-Optical and Optical-Infrared Colors}

\begin{figure}
\plotone{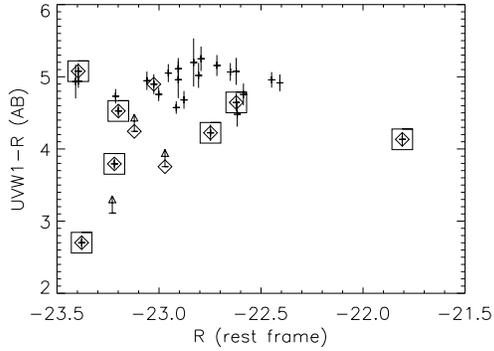}
\caption[]{UV-R AB colors in a metric aperture vs. R absolute magnitude, k-corrected to rest frame R
as in \citet{2010ApJ...713...1037H}. Diamonds are overplotted on points representing BCGs in 
cool-core clusters, as identified by Pratt et al. (2009), squares identify clusters with H$\alpha$. The
incidence of H$\alpha$ appears to be insensitive to the optical luminosity of the BCG. The colors of
BCGs lacking H$\alpha$ also appear to be insensitive to the BCG optical luminosity. \label{R}}
\end{figure}

As we showed in \S~5, 
the UVW1-R colors of the inactive galaxies
are $\sim 4.7$, with a typical reddest color of $\sim 5$. The galaxies
with emission lines comprise the outliers in the observed color relationships.
Because this sample includes brightest cluster galaxies only, 
the dynamic range in stellar mass and optical luminosity 
is not large. 
Over the luminosity range spanned by our sample, we do not detect any
trend in the observed UV-optical colors of the BCG sample with 
R magnitude or, if M/L is similar,  
stellar mass inside a $10h^{-1}$ kpc aperture (Figure~\ref{R}). Notably,  
the emission-line systems identify nearly all of the systems that differ the most
from the mean UV, optical and near-IR relations for these BCGs.

\begin{figure}
\plotone{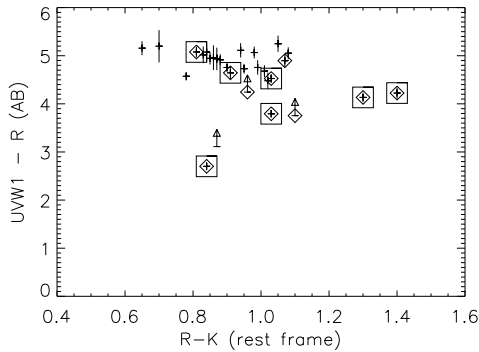}
\caption[]{UV-R AB colors in a metric aperture (this work) vs. R-K rest-frame colors from \citet{2010ApJ...713...1037H}.
Points are coded as in Figure~\ref{R}.  \label{UVRK}}
\end{figure}

We note that the scatter in the corresponding rest-frame R-K colors for the 30 galaxies in our sample
with 2MASS \citep{2006AJ....131.1163S} K-band measurements is only 0.15 ($0.95\pm0.15$), from \citet{2010ApJ...713...1037H}. 
Excluding the two BCGs with red colors (R-K$ > 1.2$), RXCJ 1302.8-0230 and RXCJ 2319.6-7313 (both
of these are emission-line BCGs in cool core clusters), the
dispersion drops to 0.11 ($0.92 \pm 0.11$), and is not much larger than the B-R scatter ($\sim0.06$)
in metric colors reported for a low redshift BCG sample 
by \citet{1995ApJ...440...28P}. A similar result is obtained if all REXCESS cool-core BCGs
are omitted from the estimation of the mean color.
The typical measurement uncertainty for R-K is about 0.10 mag, including systematics. 
Most of the statistical uncertainty arises from the 2MASS K-band photometry.
But some $\sim0.05$ mag of this uncertainty is the systematic offset between SOAR and 
2MASS photometry, and should not add to the scatter. Therefore some of the scatter in 
R-K is intrinsic, and is related to the metallicity
and age of the underlying populations. 
If R-K and UVW1-R colors were sensitive to the same underlying condition (metallicity and age) there might
be a correlation, with the UVW1-R scatter being more sensitive. 
To test whether the scatter in R-K is correlated with the scatter in UVW1-R, we plotted
these colors in Figure~\ref{UVRK}. Since we see that
UVW1-R is not correlated with R-K for the inactive BCGs, 
we can conclude that whatever variations in the stellar population
drive variations in the UV-R color (presumably metallicity and age), those variations 
are not detectable in R-K colors measured to 8\% or better. 

The BCGs in the sample with the 
reddest outlier R-K colors (RXJ 1302.8-0230 and RXJ 2319.6-7313) exhibit 
somewhat bluer UV-optical colors (a $UVW1-R \sim 4$, compared to a mean of $4.9$). They show no
obvious central point sources (indeed, no BCG in our sample does), and these two galaxies exhibit   
weak but detectable H$\alpha$ emission. The H$\alpha$ emission and a redder R-K color  may be
indicative of the contribution from a recently formed sub-population of stars. 
There are at least two counter arguments to this hypothesis. 
(1) The [NII]/H$\alpha$ ratios in these two systems are higher than seen in HII regions.
As mentioned in \S~\ref{emcol}, the high ratios in weak emission-line 
systems may indicate processes at work other than star formation. 
(2) The two most powerful emission-line BCGs (RXJ 2014.8-2430 and RXJ 1141.4-1216), 
exhibit R-K colors typical for the REXCESS sample, suggesting that excess flux from 
emission lines themselves is not sufficient to cause color anomalies. The equivalent widths of the
emission lines in the 2 R-K outliers are also very small compared to the width of the R-band filter.
Therefore, we have no ready explanation for the two outliers, aside from the fact they are notably 
BCGs with line emission, clearly not inactive.

\subsubsection{Lack of Trend of UVW1-Optical Color with Redshift or Halo Properties}

\begin{figure}
\plotone{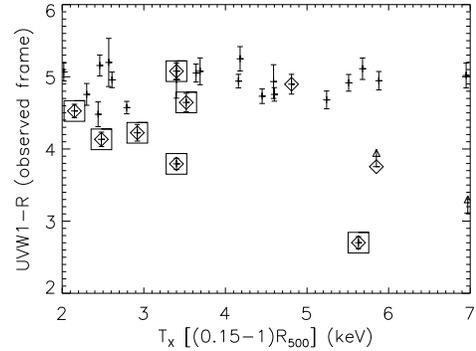}
\caption[]{UVW1-R AB colors and 3$\sigma$ lower limits plotted as a function of 
X-ray temperature. The temperature is the spectroscopic temperature from 
the region between 0.15-1 $R_{500}$ \citep{2009A&A...498..361P}. Points are coded as in Figure~\ref{R}.
We note little dependence of the UV-optical color of the BCG or the presence of H$\alpha$ emission (boxed points) from 
the BCG with X-ray temperature of the cluster. 
\label{TX}}
\end{figure}

Our data show no trend of UV-optical colors with redshift, galaxy mass, or the mass 
of the cluster of galaxies. 
Figure ~\ref{UVW1} shows that inactive BCGs in the sample exhibit no correlation of
UVW1-optical color with redshift. 
	Within the range of absolute magnitudes probed by these BCGs, we do not see a trend with optical luminosity or
R-K color. Since the optical mass-to-light ratios of these galaxies are likely to be similar, this 
lack of trend implies that the UVW1-optical colors of our inactive sample are also insensitive to the mass of the
host galaxy. 
Figure ~\ref{TX} shows no correlation of UV-optical color with X-ray 
	temperature for the inactive BCGs in our sample. Since $T_X$ is strongly correlated
	with cluster gravitating mass, this result suggests the UVW1-optical color of the BCG is insensitive to the mass of the
	host cluster.

\begin{figure}
\plotone{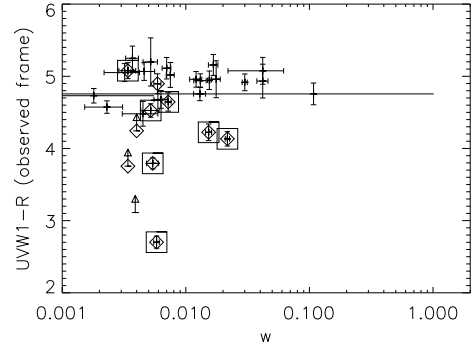}
\caption[]{UVW1-R AB colors and 3$\sigma$ lower limits plotted as a function of 
$\langle w \rangle$, which is a  standard deviation of the distance between
the X-ray peak and the X-ray emission centroid evaluated in apertures of increasing
size \citep{2009arXiv0912.4667B}. Points are coded as in Figure~\ref{R}.
We note little dependence of the UV-optical color of the BCG with this measure of cluster
dynamical state. BCGs with UV-optical colors and H$\alpha$ emission (boxed points) 
suggestive of star formation prefer to lie
in clusters with relatively low measures of  $\langle w \rangle$. The two possible exceptions to this
trend are the two BCGs with weak H$\alpha$ and unusually red R-K colors in the clusters
RXCJ1302.8-0230 and RXCJ2319.6-7313. These are also the only two clusters in REXCESS classified as both 
cool core and disturbed in Pratt et al. (2009).
\label{w}}
\end{figure}

In Figure~\ref{w} we show a plot of UV  
excess vs. $\langle w \rangle$, where the latter is the standard  
deviation of the distance between the X-ray peak and emission centroid  
evaluated in increasing apertures, as detailed in \citet{2009arXiv0912.4667B}.  
This parameter has been shown to reflect quite closely the  
true dynamical state of a system in the simulations of \citet{2006MNRAS.373..881P}.
 As can be seen in Figure~\ref{w}, there is no clear trend of UV  
excess with dynamical state. 

In summary, we can say that the lack of observed trends
suggests that no matter what the level of undetected star formation in the inactive BCGs might be, the UVW1-R  
colors of inactive BCGs  are not affected by the degree of
dynamical relaxation measured by the parameter $\langle w \rangle$, by the mass of the host cluster, 
or indeed by anything we have measured thus far.

\subsection{Potential for Insights into the UV-Upturn Phenomenon}

The bluer UV photometry measurements, including UVM2 photometry from the XMM Optical Monitor, and the
NUV (with similar but broader than the UVM2 bandpass) and FUV photometry from GALEX, are more sensitive
to the contributions of the UV-upturn population, the eHBs, than the UVW1 filter. However, for this
analysis, the data were too shallow to detect most of the BCGs in the sample and the colors even for
the detected galaxies are not very accurate, especially for the inactive galaxies.
Our primary interest in this paper is in 
the star formation properties of the BCGs and how they correlate with cluster and BCG properties, 
but we briefly discuss the potential of this sample  
for studying the properties of UV emission from older stars. 

The mechanism for creating an extreme horizontal branch star is unknown, but a survey of the literature
revealed two classes of possibilities for prematurely exposing the hot core of a horizontal branch star: 
(1) Extreme winds, sensitive to metallicity \citep[e.g.][]{2008ASPC..392....3Y} and (2) Mass transfer in 
binary stars \citep{2007MNRAS.380.1098H}. A possibly relevant wrinkle to these mechansims is based on the idea that helium 
preferentially settles in the cores of massive clusters, enhancing the UV produced by 
helium core-burning stars \citep{2009ApJ...705L..58P, 2005ApJ...621L..57L}. 

The wind-based explanations for the eHB stars predict a
 trend of reddening UV-optical colors with redshift since $z=0.2$ \citep[e.g.][]{2008ASPC..392....3Y}.
Whether variations in the relative UV component 
from galaxy to galaxy are indeed correlated with metallicity is still a matter of debate: \citet{1988ApJ...328..440B}
reported a correlation between Mg$_2$ index, considered a metallicity indicator,
and UV excess, but \citet{2005ApJ...619L.107R} see no correlation in a sample of 172 early-type galaxies.
It is possible that star formation in lenticulars dilutes this relation in a sample of early-types, 
as \citet{2007ApJS..173..597D}
see a weak correlation in ellipticals, but none in lenticulars. However, \citet{2008MNRAS.385.2097R}
show NUV-J colors of cluster red-sequence  galaxies exhibit a strong correlation with metallicity, and suggest
that a relatively tiny amount of star formation could introduce the large scatter seen even in this color.

In their GALEX study of a dozen BCGs at $z<0.2$, Ree et al (2007) report evolution of 
GALEX FUV-V colors. In contrast, the inactive BCGs (plotted without diamonds or squares
in Figure~\ref{FUV})
in our sample do not differ in their UV-optical or FUV-optical color in a significant 
way from each other or from the three low redshift template
galaxies, one of which (NGC1399) is the bluest local galaxy in Ree et al. (2007). 
Assuming $V-R \sim 0.5$, the 11 $FUV-R$ measurements in our sample are somewhat bluer than 
the dozen galaxies discussed 
in Ree et al. (2007). The REXCESS non-detections must be redder than the detections, and therefore the
colors of those galaxies likely overlap their
measurements. However, the presence of inactive but blue BCGs at $z\sim0.1$ 
suggests that the evolution may not be so strong as suggested by the single star wind models. 
Alternatively, UV emission from recently formed stars, in galaxies without detectable
H$\alpha$ emission, would confound empirical tests of this model.
The comparison to the Ree et al. (2007) dataset 
is further compromised by several limitations. Ree et al. (2007) use estimated total optical magnitudes,
while our colors are measured in fixed metric apertures. A total optical magnitude of a BCG
is difficult to measure definitively, because they often have an
extended halo of stars that fades into intracluster light, whose luminosity is comparable
to that of the central portions of the BCG \citep{2005ApJ...618..195G}. 
The UV emission seems more compact than the optical emission,
and so the choice of aperture size affects the inferred color. 
Furthermore, the shallowness of most of the archival FUV observations
covering our sample means we can only detect the bluest examples of the brightest
galaxies, with color uncertainties ranging from 0.1-0.4, and thus we may be detecting only the
galaxies in our sample with recent star formation.
Given those limitations of the archival data, the similarity of the colors may even be
surprising. Deeper observations with more accurate FUV fluxes and compatible optical
measurements are required for a comparison with the full REXCESS BCG sample, to probe the
range of FUV colors in this sample. Accurate FUV-NUV colors would be a useful diagnostic in distinguishing
the contribution of UV upturn stars from non-ionizing but relatively young main-sequence stars. 

The helium excess model, further developed by Peng \& Nagai (2009), 
predicts that the amplitude of the UV upturn should correlate  
with the dynamical state of a cluster. A high helium abundance affects the production and the
UV upturn strength of eHB stars \citep[e.g. ][]{1995ApJ...442..105D}.
In the model of \citet{2009ApJ...705L..58P}, the central helium abundance in a cluster of galaxies is enhanced by sedimentation 
in relaxed systems. If the UV emission in the inactive systems is dominated by these old stars (and not low
levels of star formation), the lack of correlation with dynamical state suggests 
that helium sedimentation  cannot explain the observed range of UV colors in these galaxies. 

Intriguingly, a binary star model for origin of the UV upturn \citep{2007MNRAS.380.1098H}
predicts an almost constant UV-optical color. In this model, binary interactions
strip horizontal branch stars. This process is independent of metallicity and redshift.

All of the models are difficult to challenge with observations, because tiny amounts of star formation, impossible
to detect with conventional emission line diagnostics, may obscure or induce trends, particularly
at 250 nm. It is notable, for example, that to reproduce the observed level of scatter 
in UV-optical or UV-infrared colors, \citet{2007MNRAS.380.1098H} must add a second, younger 
stellar population to produce significant amounts of excess UV in an otherwise inactive 
BCG or elliptical. Disentangling the contributions of
the various populations of these systems (commonly regarded as simple, single age, elderly populations 
in the past) requires more UV work, particularly spectroscopy, to obtain the needed
diagnostics. Deeper GALEX photometry of the REXCESS BCG sample would be useful to determine whether there are empirical trends in these high-mass objects. More sophisticated simulations are required to determine
whether trends predicted from the simpler models we have in hand 
could be detected under more realistic conditions.

\section{Summary and Conclusion}

We report the results of UV broad-band photometry from the XMM Optical Monitor and from the GALEX
mission and of long-slit optical emission-line spectroscopy for 32 BCGs in REXCESS. 
Seven of these clusters exhibit classic signatures of star formation activity associated
with a cool-core cluster as a host. Indeed these seven BCGs inhabit the  
10 clusters in the REXCESS sample with the shortest cooling times in the hot gas, as inferred
from XMM observations, and the two most luminous of these in H$\alpha$  
are the most prominent cool core clusters in REXCESS. The 
incidence rate of emission-line BCGs is intermediate between that found in the B55 or
EMSS cluster samples and that found in optically-selected SDSS clusters, possibly a
consequence of the lack of morphological bias in the selection of REXCESS clusters. The BCGs with
the largest H$\alpha$ equivalent widths are also the BCGs with the bluest UV-optical colors.
We report a correlation between the BCG H$\alpha$ equivalent width and -- to a lesser degree -- the
UV-optical color with conditions in the intracluster gas in the cluster core: the scaled core 
electron density and the gas cooling time. The incidence rates and the correlations suggest 
a physical connection between activity (emission-line excitation, recent star formation) 
in the BCG and the cooling time of the intracluster gas.   
We see no correlation between H$\alpha$ equivalent
width and X-ray temperature or cluster mass; we also see no correlation between the UV-optical
color and cluster mass, temperature, or degree of relaxation, suggesting that whatever ignites BCG activity at
low redshift is relatively insensitive to halo properties. This insensitivity is puzzling, given
the difference in the incidence of emission-line BCGs in optically selected vs. X-ray
selected samples of clusters of galaxies. A decreasing incidence (or strength) of emission-line BCGs
with decreasing halo mass or BCG mass could have explained the discrepancy. However, we do not
have evidence for a correlation of that nature, at least over the range probed by the REXCESS cluster
sample. However, the strong correlation between the presence of low entropy gas and the 
appearance of star formation signatures suggests there is a connection between the nature of
the cluster and conditions in its BCG. These star-forming 
BCGs also tend to inhabit clusters with low $\langle w \rangle$,
characteristic of clusters with relatively relaxed dynamical states. Optically selected cluster samples 
may include a wider range of morphologies and dynamical states than that sampled by X-ray selected
cluster samples.

Almost all ($29/31$) of the BCGs were detected in the OM UVW1 band, regardless of their
emission-line activity. The UVW1-optical colors of these galaxies are consistent with the UV light
of an old population, with excess UV coming from recent
star formation. The UVW1 photometry is not strongly affected by the UV upturn stars, and 
therefore turns out to be a decent choice to study UV associated with star formation. We analyze
archival GALEX and XMM OM UVM2 data at shorter wavelengths.

The inactive BCGs in our sample, classified as such by the lack of H$\alpha$ emission, define a relatively
homogeneous sample.  The observed UV-optical colors of these BCGs (UVW1-R$ \sim 4.8$) are 
independent of redshift, BCG absolute rest frame R magnitude, and rest frame R-K color, suggesting that 
the UVW1-optical colors of inactive BCGs are also unrelated to BCG mass and metallicity. Their
colors are independent of cluster temperature (a measure of the depth of the cluster's 
gravitational potential) and degree of relaxation, as measured by the offset of the X-ray centroid from
the X-ray emission peak. This lack of correlation means that the UVW1-R colors are not affected
by the gravitating mass of the cluster or its dynamical state.

\acknowledgements
We acknowledge the work of Susanne Heidenreich for providing the radio information for these 
galaxies in advance of publication. We are grateful to 
Thomas M. Brown for his digital version of the three UV upturn elliptical galaxy spectral templates.
MD, AH, GMV, and EW received significant support from a NASA Long Term Space Astrophysics grant (NNG-05GD82G).
The present work is  
partly based on observations obtained with {\it XMM-Newton}, an ESA  
science mission with instruments and contributions directly funded by  
ESA Member States and the USA (NASA). 
The SOAR Telescope is a joint project of Conselho Nacional des Pesquisas Cientficas e Tecnologicas 
CNPq-Brazil, The University of North Carolina Chapel Hill, Michigan State University, and the National Optical
Astronomy Observatory. This research has made use of NASA's Astrophysics Data System, the Multimission
Archive at Space Telescope Science Institute, and the NASA Goddard High Energy Astrophysics Science 
Archive Research Center. MD also warmly acknowledges support from Carnegie Observatories and the Las Campanas
Observatory staff and telescope operators, from long ago but not forgotten.

\bibliography{rexcess}

\clearpage
\begin{deluxetable}{lccccc}
\tablecaption{XMM Optical Monitor Observations}
\tablewidth{0pt}
\tablecolumns{6}
\tablehead{\colhead{Target} & \colhead{Observation ID} &\colhead{Filter} & \colhead{Mode} & \colhead{Exposure Time} & \colhead{Observation Date} \\
\colhead{2MASS ID} & \colhead{}  & \colhead{} & \colhead{} & \colhead{(seconds)}& \colhead{(YYYY-MM-DD)} }
\startdata
00034964+0203594 & 0201900101 & uvm2 & HR & 9900 & 2004 Jun 24\\
00055975-3443171 & 0201903801 & uvm2 & HR & 12500 & 2005 May 13\\
00204314-2542284 & 0201900301 & uvm2 & HR & 12300 & 2004 May 26\\
00492282-2931069 & 0201900401 & uvm2 & HR & 16700 & 2004 Dec 04\\
01445891-5301110 & 0201900501 & uvm2 & HR & 15200 & 2004 Nov 12\\
02112484-4017261 & 0201900601 & uvm2 & HR & 13300 & 2004 Dec 27\\
02250904-2928383 & 0201900701 & uvm2 & HR & 11400 & 2004 Jul 06\\
05473773-3152237 & 0201900901 & uvm2 & HR & 10900 & 2004 Mar 07\\
06055401-3518081 & 0201901001 & uvm2 & LR & 4960 & 2004 Oct 29\\
06165166-4747434 & 0201901101 & uvm2 & HR & 21800 & 2004 Apr 26\\
06452948-5413365 & 0201903401 & uvm2 & HR & 8700 & 2004 Jun 12\\
08215065+0111495 & 0201901301 & uvm2 & LR & 2840 & 2004 Oct 13\\
09582201-1103500 & 0201903501 & uvm2 & HR & 5500 & 2004 Jun 17\\
10443287-0704074 & 0201901501 & uvm2 & HR & 9360 & 2004 Dec 24\\
11412420-1216386 & 0201901601 & uvm2 & HR & 17200 & 2004 Jul 09\\
12364125-3355321 & 0201901701 & uvm2 & LR & 2880 & 2004 Jul 28\\
13025254-0230590 & 0201901801 & uvm2 & HR & 9400 & 2004 Jun 22\\
15161794+0005203 & 0201902001 & uvm2 & HR & 12700 & 2004 Jul 23\\
15164416-0058096 & 0201902101 & uvm2 & LR & 2320 & 2004 Aug 03\\
20145171-2430229 & 0201902201 & uvm2 & HR & 9900 & 2004 Oct 08\\
20225911-2056561 & 0201902301 & uvm2 & HR & 13400 & 2005 Apr 06\\
20481162-1749034 & 0201902401 & uvm2 & LR & 1980 & 2004 May 13\\
21490737-3042043 & 0201902601 & uvm2 & HR & 9360 & 2004 Nov 29\\
21520957-1943235 & 0201902701 & uvm2 & LR & 2340 & 2004 Oct 28\\
21572939-0747443 & 0201902801 & uvm2 & LR & 2340 & 2005 May 11\\
22174585-3543293 & 0201902901 & uvm2 & LR & 2340 & 2005 May 12\\
22183938-3854018 & 0201903001 & uvm2 & HR & 10700 & 2004 Oct 24\\
22342463-3743304 & 0201903101 & uvm2 & LR & 2720 & 2004 Nov 11\\
23194046-7313366 & 0201903201 & uvm2 & LR & 2340 & 2004 Apr 18\\ \tableline \tableline
00034964+0203594 & 0201900101 & uvw1 & HR & 9900 & 2004 Jun 24\\
00055975-3443171 & 0201900201 & uvw1 & HR & 12160 & 2004 Dec 08\\
00204314-2542284 & 0201900301 & uvw1 & HR & 12300 & 2004 May 26\\
00492282-2931069 & 0201900401 & uvw1 & HR & 16700 & 2004 Dec 04\\
01445891-5301110 & 0201900501 & uvw1 & HR & 16200 & 2004 Nov 12\\
02112484-4017261 & 0201900601 & uvw1 & HR & 12500 & 2004 Dec 27\\
02250904-2928383\tablenotemark{a} & 0201900701 & uvw1 & HR & 11400 & 2004 Jul 06\\
02250904-2928383\tablenotemark{a} & 0302610601 & uvw1 & HR & 9700 & 2006 Jan 27\\
05473773-3152237 & 0201900901 & uvw1 & HR & 10900 & 2004 Mar 07\\
06055401-3518081 & 0201901001 & uvw1 & HR & 11800 & 2004 Oct 29\\
06165166-4747434 & 0201901101 & uvw1 & HR & 16900 & 2004 Apr 26\\
08215065+0111495 & 0201901301 & uvw1 & LR & 2240 & 2004 Dec 13\\
09582201-1103500 & 0201903501 & uvw1 & LR & 820 & 2004 Jun 17\\
10443287-0704074 & 0201901501 & uvw1 & HR & 12400 & 2004 Dec 23\\
11412420-1216386 & 0201901601 & uvw1 & HR & 13760 & 2004 Jul 09\\
12364125-3355321 & 0201903701 & uvw1 & LR & 1220 & 2004 Dec 30\\
13025254-0230590 & 0201901801 & uvw1 & HR & 1880 & 2004 Jun 22\\
13112952-0120280 & 0093030101 & uvw1 & HR & 6300 & 2001 Dec 24\\
15161794+0005203 & 0201902001 & uvw1 & HR & 12200 & 2004 Jul 22\\
15164416-0058096 & 0201902101 & uvw1 & LR & 2320 & 2004 Aug 03\\
20145171-2430229 & 0201902201 & uvw1 & HR & 9900 & 2004 Oct 08\\
20225911-2056561 & 0201902301 & uvw1 & HR & 7740 & 2005 Apr 06\\
20481162-1749034 & 0201902401 & uvw1 & LR & 1980 & 2004 May 13\\
21294244-5049260 & 0201902501 & uvw1 & LR & 2340 & 2004 Oct 16\\
21490737-3042043 & 0201902601 & uvw1 & HR & 11700 & 2004 Nov 29\\
21520957-1943235 & 0201902701 & uvw1 & LR & 2340 & 2004 Dec 28\\
21572939-0747443 & 0201902801 & uvw1 & LR & 2340 & 2005 May 11\\
22174585-3543293 & 0201902901 & uvw1 & HR & 11700 & 2005 May 12\\
22183938-3854018 & 0201903001 & uvw1 & HR & 10700 & 2004 Oct 24\\
22342463-3743304 & 0018741701 & uvw1 & HR & 5000 & 2001 May 03\\
23194046-7313366 & 0201903201 & uvw1 & HR & 11700 & 2004 Apr 18\\

\enddata
\label{table:om}
\tablenotetext{a}{AB magnitude was obtained by taking a weighted average of the two exposures.}
\end{deluxetable}

\begin{deluxetable}{llcccccccc}
\footnotesize
\tablecaption{Metric Aperture UV Optical Monitor Photometry}
\tablecolumns{10}
\tablewidth{0pt}
\tablehead{\colhead{Cluster} & \colhead{2MASS ID\tablenotemark{b}} &  \colhead{z} & \colhead{r} & \colhead{UVW1} & \colhead{UVW1} & \colhead{UVM2} & \colhead{UVM2} & \colhead{R} & \colhead{$A_V$} \\ 
\colhead{} & \colhead{} &  \colhead{} &\colhead{$\arcsec$} & \colhead{mag} & \colhead{Unc} & \colhead{mag} & \colhead{Unc} & \colhead{mag} & \colhead{}} 

\startdata
RXCJ0003.8+0203 & 00034964+0203594 & 0.092 & 8.3 & 20.54 & 0.10 & 21.61 & 0.00 & 15.35 & 0.11 \\ 
RXCJ0006.0-3443 & 00055975-3443171 & 0.115 & 7.0 & 20.64 & 0.06 & 22.13 & 0.35 & 15.80 & 0.061 \\ 
RXCJ0020.7-2542 & 00204314-2542284 & 0.141 & 5.8 & 21.18 & 0.10 & 22.21 & 0.00 & 16.44 & 0.052 \\ 
RXCJ0049.4-2931 & 00492286-2931124 & 0.108 & 7.2 & 20.42 & 0.05 & 22.04 & 0.00 & 15.75 & 0.073 \\ 
RXCJ0145.0-5300 & 01445891-5301110 & 0.117 & 6.8 & 21.53 & 0.09 & 21.93 & 0.00 & 16.48 & 0.104 \\ 
RXCJ0211.4-4017 & 02112484-4017261 & 0.101 & 7.7 & 20.98 & 0.10 & 21.86 & 0.31 & 15.84 & 0.06 \\ 
RXCJ0225.1-2928 & 02250904-2928383 & 0.06 & 12.3 & 19.84 & 0.08 & 21.31 & 0.00 & 14.81 & 0.06 \\ 
RXCJ0345.7-4112 & 03454640-4112149 & 0.06 & 0.0 & 18.63 & 0.06 & 20.2 & 0.3 & 14.05 & 0.046 \\ 
RXCJ0547.6-3152 & 05473773-3152237 & 0.148 & 5.5 & 21.81 & 0.13 & 22.08 & 0.00 & 16.57 & 0.098 \\ 
RXCJ0605.8-3518 & 06055401-3518081 & 0.139 & 5.8 & 21.53 & 0.12 & 20.45 & 0.00 & 16.37 & 0.2 \\ 
RXCJ0616.8-4748 & 06165166-4747434 & 0.116 & 6.8 & 20.68 & 0.06 & 22.23 & 0.00 & 15.53 & 0.169 \\ 
RXCJ0645.4-5413 & 06452948-5413365 & 0.164 & 5.1 & 20.2 & 0.00 & 22.00 & 0.00 & 16.69 & 0.312 \\ 
RXCJ0821.8+0112 & 08215065+0111495 & 0.082 & 9.2 & 20.09 & 0.16 & 21.04 & 0.00 & 15.43 & 0.139 \\ 
RXCJ0958.3-1103 & 09582201-1103500 & 0.167 & 5.0 & 20.94 & 0.00 & 21.64 & 0.00 & 16.90 & 0.214 \\ 
RXCJ1044.5-0704 & 10443287-0704074 & 0.134 & 6.0 & 21.43 & 0.11 & 21.75 & 0.00 & 16.63 & 0.122 \\ 
RXCJ1141.4-1216 & 11412420-1216386 & 0.119 & 6.6 & 19.64 & 0.03 & 20.01 & 0.05 & 15.72 & 0.102 \\ 
RXCJ1236.7-3354 & 12364125-3355321 & 0.08 & 9.5 & 20.77 & 0.33 & 21.07 & 0.00 & 15.25 & 0.251 \\ 
RXCJ1302.8-0230 & 13025254-0230590 & 0.085 & 9.0 & 19.67 & 0.09 & 20.99 & 0.27 & 15.34 & 0.078 \\ 
RXCJ1311.4-0120 & 13112952-0120280 & 0.183 & 4.6 & 21.25 & 0.00 & 0.00 & 0.00 & 16.89 & 0.088 \\ 
RXCJ1516.3+0005 & 15161794+0005203 & 0.118 & 6.7 & 21.67 & 0.15 & 21.98 & 0.00 & 16.18 & 0.183 \\ 
RXCJ1516.5-0056 & 15164416-0058096 & 0.12 & 6.6 & 21.41 & 0.25 & 21.21 & 0.00 & 16.15 & 0.23 \\ 
RXCJ2014.8-2430 & 20145171-2430229 & 0.155\tablenotemark{a} & 5.4 & 19.96 & 0.04 & 20.51 & 0.10 & 16.62 & 0.494 \\ 
RXCJ2023.0-2056 & 20225911-2056561 & 0.056 & 13.1 & 19.89 & 0.13 & 21.07 & 0.00 & 14.48 & 0.192 \\ 
RXCJ2048.1-1750 & 20481162-1749034 & 0.147 & 5.5 & 21.32 & 0.22 & 20.81 & 0.00 & 16.13 & 0.199 \\ 
RXCJ2129.8-5048 & 21294244-5049260 & 0.0796 & 9.5 & 20.49 & 0.17 & 0.00 & 0.00 & 15.31 & 0.085 \\ 
RXCJ2149.1-3041 & 21490737-3042043 & 0.118 & 6.7 & 20.71 & 0.09 & 21.80 & 0.00 & 15.51 & 0.093 \\ 
RXCJ2152.2-1942 & 21520957-1943235 & 0.096 & 8.0 & 20.10 & 0.10 & 20.32 & 0.00 & \nodata  & 0.107 \\ 
RXCJ2157.4-0747 & 21572939-0747443 & 0.058 & 12.7 & 19.56 & 0.13 & 20.48 & 0.00 & 14.64 & 0.129 \\ 
RXCJ2217.7-3543 & 22174585-3543293 & 0.149 & 5.5 & 21.06 & 0.07 & 21.86 & 0.24 & 16.25 & 0.059 \\ 
RXCJ2218.6-3853 & 22183938-3854018 & 0.141 & 5.8 & 21.26 & 0.11 & 22.06 & 0.00 & 16.25 & 0.05 \\ 
RXCJ2234.5-3744 & 22342463-3743304 & 0.151 & 5.4 & 21.77 & 0.16 & 21.34 & 0.00 & 16.68 & 0.057 \\ 
RXCJ2234.5-3744b & 22342463-3743304 & 0.151 & 5.4 & 20.59 & 0.07 & 22.10 & 0.33 & \nodata & 0.057 \\ 
RXCJ2319.6-7313 & 23194046-7313366 & 0.098 & 7.9 & 20.91 & 0.07 & 21.81 & 0.30 & 16.64 & 0.101 \\ 

\enddata
\tablecomments{All magnitudes are in the AB system and measured within an aperture of radius $r=10 h^{-1}=14.3$ kpc, reported in column 4, 
along with a flat, $\Omega_M=0.3$ geometry. The R band magnitudes have a systematic uncertainty of $\sim0.05$. The 
magnitudes listed are not corrected for Galactic extinction. No internal extinction is applied. 
The $A_V$ in the direction of the BCG listed, and we used this to correct the UV and optical colors. All colors in
the plots have been corrected for Galactic extinction.}
\tablenotetext{a}{The X-ray spectra were incompatible with the NED redshift for RXCJ2014.8-2430 of 0.16. Our optical
spectroscopy reported in this work indicates a redshift of 0.1555, consistent with the redshift estimated from the 
X-ray spectrum of 0.1538 from Pratt et al. (2009). }
\tablenotetext{b}{All of these IDs are from the 2MASS Extended Source Catalog except for 02250904-2928383 and
00492282-2931069, which were compact and round enough in 2MASS observations for the Point Source Catalog.}
\label{table:omphot}
\end{deluxetable}

\begin{deluxetable}{lcccc|cccc|rr}
\footnotesize
\tablecaption{GALEX Photometry}
\tablecolumns{11}
\tablewidth{0pt}
\tablehead{
	\multicolumn{1}{c}{} & \multicolumn{4}{c|}{Matched Aperture (r=14 kpc)} & \multicolumn{4}{c|}{Total Flux} & \multicolumn{2}{c}{Exposure Time} \\[4pt] 
	\colhead{Cluster} & \colhead{NUV} & \colhead{NUV} & \colhead{FUV} & \colhead{FUV} & \colhead{NUV} & \colhead{NUV} & \colhead{FUV} & \colhead{FUV} & \colhead{NUV} & \colhead{FUV} \\
	\colhead{} & \colhead{AB} & \colhead{Unc}  & \colhead{AB} & \colhead{Unc}& \colhead{AB} & \colhead{Unc}& \colhead{AB} & \colhead{Unc} & \colhead{(s)} & \colhead{(s)}} \\

\startdata
RXCJ0003.8+0203 & 22.0  & 0.3  & 21.6  & 0.5 & 21.1  & 0.3  & 21.5  & 0.4 & 112 & 342 \\
RXCJ0006.0-3443 & 22.0  & 0.4  & 21.85 & 0.85 & 21.12 & 0.28 & 21.57 & 0.22 & 1608 & 1608 \\ 
RXCJ0020.7-2542 & 22.7  & 0.4  & 22.1  & 0.4 & 21.7  & 0.4  & 22.4  & 0.4 & 325 & 325 \\
RXCJ0049.4-2931 & 20.93 & 0.16  & \nodata & \nodata  & 20.22 & 0.13 & $<21.6$ & \nodata  & 203 & 203 \\
RXCJ0225.1-2928 & 20.75 & 0.20  & \nodata & \nodata  & 20.04 & 0.23 & $<21.5$ & \nodata & 169 & 169 \\
RXCJ0345.7-4112 & 19.85 & 0.14  & 20.50 & 0.27 & 19.40 & 0.15 & 20.5 & 0.3 & 106 & 106 \\
RXCJ0821.8+0112\tablenotemark{a} & 21.9  & 0.5   & \nodata & \nodata  & 20.0 & 0.4 & $<21.5$ & \nodata & 173 & 168 \\
RXCJ1141.4-1216 & 19.85 & 0.09  & 20.19 & 0.30 & 19.68 & 0.09 & 19.74 & 0.24 & 186 & 75 \\
RXCJ1302.8-0230 & 20.99 & 0.06  & 21.59 & 0.11 & 20.31 & 0.08 & 21.34 & 0.12 & 1700 & 1700 \\
RXCJ2014.8-2430\tablenotemark{b} & 20.20 & 0.08  &  \nodata & \nodata & 19.84 & 0.085 & \nodata & \nodata & 365 & \nodata \\ 
RXCJ2023.0-2056 & 20.9 & 0.3    & 21.0   & 0.4   & 21.1 & 0.3 & 20.7 & 0.3 & 184 & 184 \\
RXCJ2129.8-5048\tablenotemark{c} & 21.3 & 0.3    & \nodata & \nodata  & 21   & 0.3 & $<21.6$ & \nodata & 207  & 207 \\
RXCJ2157.4-0747 & 20.73 & 0.08  & 22.00 & 0.17 & 20.65 & 0.06 & 21.7 & 0.11 & 2973 & 2973 \\
RXCJ2217.7-3543 & 22.0  & 0.2   & 22.0  & 0.3  & 21.   & 0.2  & 21.7 & 0.3  & 404 & 311 \\
RXCJ2218.6-3853\tablenotemark{a} & 22.4  & 0.4   & \nodata & \nodata  & 22.2 & 0.4 & $<21.5$ & \nodata & 180 & 180 \\
RXCJ2234.5-3744ab\tablenotemark{d} & 21.04 & 0.01 & 21.37 & 0.02 & 20.57 & 0.01 & 21.04 & 0.02 & 27789 & 27789 \\
RXCJ2319.6-7313 & 21.7 & 0.5 & \nodata & \nodata  & 21.3 & 0.4 & $<21.2$ & \nodata & 108 & 108 \\
\enddata

\tablecomments{Metric aperture ($10h^{-1}$ kpc) photometry is interpolated to match the apertures reported in Table~\ref{table:omphot}. No GALEX data were available for RXCJ0145, RXCJ0547, RXCJ0605, RXCJ0645, RXCJ1311, RXCJ1516.3, RXCJ2149, or RXCJ2152. The seven remaining
undetected BCGs have only upper limits ($\lesssim 21$ in FUV, NUV: RXCJ0211, RXCJ0616, RXCJ0958, RXCJ1044, RXCJ1236, RXCJ1516.5, RXCJ2048.  }
\tablenotetext{a}{The fluxes for these galaxies are included in the table for completeness, but are highly uncertain, and 
represent the lowest confidence GALEX detections in the sample.}
\tablenotetext{b}{No FUV data were available for RXCJ2014.}
\tablenotetext{c}{This GALEX source is off-center by $8\arcsec$, but is within the BCG.}
\tablenotetext{d}{The BCG for RXCJ2234 is included in a blended GALEX source. Higher resolution OM data show two relatively bright components and a faint component. The bright source is unlikely to be associated with the BCG.}
\label{table:galexphot}
\end{deluxetable}

\begin{deluxetable}{lccccc}
\tablecaption{SOAR Observations}
\tablewidth{0pt}
\tablecolumns{6}
\tablehead{
\colhead{Target} & \colhead{Observation} & \colhead{Exposure}&\colhead{St. Star}&\colhead{Position Angles}&\colhead{H$\alpha$ EQW}\\
\colhead{} & \colhead{Date}  & \colhead{(seconds)} & \colhead{} & \colhead{(degrees E of N)} & \colhead{$-$(\AA)}}

\startdata

00034964+0203594 & 2008 Oct 05  &  1200 & LTT7379 & 75, 165 & $<1.4$ \\
00055975-3443171 & 2008 Nov 02\tablenotemark{a}  & 1500 & LTT377 & 15, 105   & $<1.4$ \\
00204314-2542284 & 2008 Nov 26  & 1500 & LTT1020 & 20, 110   & $<2.2$ \\
00492286-2931124 & 2008 Nov 26  & 1500 &  LTT1020 & 0, 90     & $<1.6$ \\
01445891-5301110 & 2008 Nov 26  & 1500 &  LTT1020 & 0, 90     & $<1.4$ \\
02112484-4017261 & 2008 Oct 04  & 1500  & LTT377 &  30       & $<1.4$ \\
02250904-2928383 & 2008 Oct 04  & 1200 & LTT377 & 5, 95     & $<0.8$ \\
03454640-4112149 & 2008 Sep 30 & 1200 & LTT9239 & 133, 43     & $6.4\pm0.5$ \\
05473773-3152237 & 2008 Sep 30 & 1200 &  LTT9239 & 111        & $<1.8$ \\
06055401-3518081 & 2008 Oct 04 &  1200 & LTT377 & 110,200   & $<2$ \\
06165166-4747434 & 2008 Nov 02\tablenotemark{a} & 1500 &LTT377 &  0          & $<1.2$ \\
06452948-5413365 & 2008 Nov 02\tablenotemark{a}  & 1500 & LTT377 & 65, 155   & $<2.4$ \\
08215065+0111495 & 2008 Nov 26  & 1500 &  LTT1020 & 75        & $<0.8$\\
09582201-1103500 & 2009 Apr 17\tablenotemark{a}   & 1500 & LTT3864 & 65, 155 & $<2.4$\\ 
10443287-0704074 & 2009 Apr 17\tablenotemark{a} &  1500 & LTT3864 & 85, 175   & $7.6\pm0.9$\\
11412420-1216386 & 2009 Apr 17\tablenotemark{a} &  1500 & LTT3864 & 70, 160 & $42.4\pm1.5$ \\
12364125-3355321 & 2009 Apr 27\tablenotemark{a} &  $500\times3$ & LTT3864 & 50& $<1$ \\
13025254-0230590 & 2009 Apr 27\tablenotemark{a} &  $500\times3$ & LTT3864 & 85, 175& $2.2\pm1.1$ \\
13112952-0120280 & \nodata & \nodata & \nodata &  & $<0.6$\tablenotemark{b} \\
15161794+0005203 & 2009 Apr 17\tablenotemark{a} & 1500 &  LTT3864 &  45, 135 & $<1.6$ \\
15164416-005809 & 2009 Apr 27\tablenotemark{a} & 1500 &  LTT3864 &  50, 140 & $<2.2$ \\
20145171-2430229 & 2009 Sep 30 & 900 &  LTT9239 &40, 130 & $73\pm2$ \\
20225911-2056561 & 2008 Oct 04 & 1200 & LTT377 & 20, 110 & $1.8$ \\
20481162-1749034 & 2008 Nov 02\tablenotemark{a} & 1500 &LTT377 &  0, 90 & $<1$ \\
21294244-5049260 & 2008 Oct 05 & 1200 & LTT7379 & 14, 104 & $<1.2$  \\
21490737-3042043 & 2008 Oct 04  & 1200 & LTT9239 & 20,110 & $3.8\pm0.5$ \\
21520957-1943235 & 2008 Sep 30  & 900 & LTT9239 & 9, 99 & $<1.4$ \\
21572939-0747443 & 2008 Nov 26 & 1500 &  LTT1020 & 170, 260 & $<2$\tablenotemark{c} \\
22174585-3543293 & 2008 Sep 30 & 1200 &  LTT9239 &78, 168 & $<3$ \\
22183938-3854018 & 2008 Sep 30 & 900 &  LTT9239 &62, 152 & $<2.2$ \\
22342463-3743304 & 2008 Oct 05  & 1200 & LTT7379 & 85, 355 & $<1.6$ \\
23194046-7313366 & 2008 Nov 02\tablenotemark{a} & 1500 & LTT377 &  85 & $3.8\pm0.5$ \\

\enddata
\tablecomments{An Exposure Time of ``$500\times3$" indicates 3 images of length 500 seconds were taken at each position angle.}
\label{table:goodman}
\tablenotetext{a}{The blue blocking filter GG-495 was used for these observations.}
\tablenotetext{b}{Based on a $3\sigma$ H$\alpha$ luminosity upper limit of $3.7 \times10^{40}$ erg s$^{-1}$ 
for the BCG in  Abell 1689, based on Las Campanas Modular Spectrograph, in June 1993. See Table~\ref{table:lco}.}
\tablenotetext{c}{Spectroscopic fringing severely limited the sensitivity of this observation.}
\end{deluxetable}

\begin{deluxetable}{lcccccccc}
\tablecaption{Emission Line Properties}
\tablewidth{0pt}
\tablecolumns{9}
\tablehead{
\colhead{BCG} & \colhead{z} & \colhead{EQW (H$\alpha$) }&\colhead{NII6584/H$\alpha$} & \colhead{FWHM} & \colhead{No. Lines} & \colhead{$L(H\alpha)$} & \colhead{SFR (H$\alpha$)} & \colhead{SFR (UV)} \\
\colhead{} & \colhead{}  & \colhead{(\AA)} & \colhead{} & \colhead{(km s$^{-1}$)} & \colhead{(No. spectra)} & \colhead{$10^{40}~h_{70}^{-2}$ erg s$^{-1}$} & \colhead{M$_\odot$ yr$^{-1}$} & \colhead{M$_\odot$ yr$^{-1}$} \\}
\startdata
RXCJ0345.7-4112 & $0.061\pm0.002$  & $-6.4\pm0.8$  & $1.9\pm0.2$ & $14.9\pm0.6$ & 5 (2) & 5.3 & 0.4 & 0.24-1.0 \\
RXCJ1044.5-0704 & $0.1330\pm0.0004$ & $-7.6\pm0.8$ & $1.4\pm0.3$ & $12.4\pm0.8$ & 3 (2) & 5.4 & 0.4 & 0.1-0.4 \\
RXCJ1141.4-1216 & $0.1190\pm0.0005$ & $-42\pm1$ & $1.38\pm0.03$ & $19.8\pm0.3$ & 6 (2)  & 35 & 2.8 & 2-4 \\
RXCJ1302.8-0230 & $0.0847\pm0.0001$ & $-2.2\pm0.9$ & $3.7\pm0.9$ & $13.7\pm2.4$ & 5 (2) & 2.5 & 0.2 & 0.5-1.2 \\
RXCJ2014.8-2430 & $0.1555\pm0.0003$ & $-73.0\pm1.3$ & $0.77\pm0.04$\tablenotemark{*} & $13.2\pm0.2$ & 6 (2) & 64 & 5.0 & 8-14  \\
RXCJ2149.1-3041 & $0.1209\pm0.0001$ & $-3.8\pm0.7$ & $2.1\pm0.4$ & $10.8\pm0.6$ & 5 (2) & 2.0 & 0.16  & 0 \\
RXCJ2319.6-7313 & $0.0981\pm0.004$ & $-10.8\pm1.4$ & $1.2\pm0.1$ & $12.3\pm0.5$ & 5 (1) & 1.6 & 0.13 & 0.3-0.6 \\
\enddata
\tablecomments{Equivalent width and FWHM are reported  for the H$\alpha$ line in the observer frame. Column 6 lists
the number of emission lines detected, and the number in parenthesis reports the number of distinct spectra (different
position angles) used. If two spectra were available, we averaged the emission line properties. Column 7 is an estimate
of the H$\alpha$ luminosity, based on the maximum emission line flux in the slit, corrected for Galactic extinction. 
This estimate is a lower limit, since the emission line flux is
usually more extended than the slit width and no correction is applied for intrinsic absorption. 
Column 8 is the star formation rate based on this luminosity \citep{1998ARA&A..36..189K}. Column 9 is
the star formation rate based on an estimate of the UV excess, uncorrected for intrinsic absorption, 
with a baseline color of UVW1-R ranging between 4.7-5 and for starbursts
ranging in ages from $10^7$ to $10^8$ years (the former giving the highest rates) \citep{1998ARA&A..36..189K, 2003A&A...410...83H}. }
\tablenotetext{*}{In this case, the [NII]6583 line was attenuated by an atmospheric absorption line, so the [NII]6583/H$\alpha$ ratio 
for RXCJ2014 is based on the measurement of the [NII]6548 line flux, which is 1/3 [NII]6583 from atomic physics.}
\label{table:emissionlines}
\end{deluxetable}

\begin{deluxetable}{lcccc}
\tablecaption{Las Campanas Modular Spectrograph Observations}
\tablewidth{0pt}
\tablecolumns{5}
\tablehead{
\colhead{Target} & \colhead{Observation} & \colhead{Exposure}&\colhead{Position Angles}&\colhead{H$\alpha$ EQW}\\
\colhead{} & \colhead{Date}  & \colhead{(seconds)} & \colhead{(degrees E of N)} & \colhead{$-$(\AA)}}
\startdata
Abell 1650  & 1993 Jun 23 & 900, 1800 & 90, 0 & $<0.6$ \\
Abell 1689  & 1993 Jun 23 & 1200, 900 & 90, 18 & $<1.1$       \\
Abell 3571  & 1992 Apr 11/12 & $3\times 900$ & 90,0 & $<0.5$   \\
TriAus      & 1992 Apr 11/12 & $2\times 900$ & 90 & $<0.5$     \\
Abell 4038\tablenotemark{a} & 1992 Dec 17 & $2\times 900$ & 117, 27 & $<1$   \\
\enddata
\tablenotetext{a}{Also known as Klemola 44.}
\label{table:lco}
\end{deluxetable}

\end{document}